\documentclass[final,3p,times,twocolumn]{elsarticle}

\usepackage{graphics}
\usepackage{graphicx}
\usepackage{epsfig}
\usepackage{amsthm}
\usepackage{amsmath}
\usepackage{amssymb}
\usepackage{lineno}
\usepackage{fleqn}
\usepackage{txfonts}
\usepackage{multirow}
\usepackage{array}

\begin{document}

\begin{frontmatter}

\title{Micromagnetic simulation of magnetic small-angle neutron scattering \\ from two-phase nanocomposites}

\author[a]{Andreas Michels\corref{cor1}}
\cortext[cor1]{Corresponding author}\ead{andreas.michels@uni.lu}
\address[a]{Laboratory for the Physics of Advanced Materials, University of Luxembourg, 162A~Avenue de la Fa\"iencerie, L-1511 Luxembourg, Luxembourg}
\author[b]{Sergey Erokhin}
\author[b]{Dmitry Berkov}
\author[b]{Nataliya Gorn}
\address[b]{INNOVENT Technology Development, Pr\"ussingstra{\ss}e 27B, D-07745 Jena, Germany}

\begin{abstract}
The recent development of a micromagnetic simulation methodology---suitable for multiphase magnetic nanocomposites---permits the computation of the magnetic microstructure and of the associated magnetic small-angle neutron scattering (SANS) cross section of these materials. In this review article we summarize results on the micromagnetic simulation of magnetic SANS from two-phase nanocomposites. The decisive advantage of this approach resides in the possibility to srutinize the individual magnetization Fourier contributions to the total magnetic SANS cross section, rather than their sum, which is generally obtained from experiment. The procedure furnishes unique and fundamental information regarding magnetic neutron scattering from nanomagnets.
\end{abstract}


\end{frontmatter}

%\linenumbers
\biboptions{sort&compress}

\section{Introduction}

Small-angle neutron scattering (SANS) is one of the most important techniques for microstructure determination in soft and hard condensed matter, materials science, and in physical chemistry. Since most SANS studies focus on nuclear rather than magnetic scattering, it is not surprising that the theoretical concepts behind nuclear SANS are rather well developped \cite{fournet,glatter,feigin,linchen,zemb,pedersen02,svergun03,stuhrmann04,wignall07}. By contrast, the understanding of magnetic SANS is still at its beginning, although magnetic SANS has previously demonstrated great potential for resolving the spin structures of various magnetic materials. For instance (in the last decade), magnetic SANS has been employed for studying the microstructures of magnetic nanocomposites \cite{heinemann2000,hermann2000b,michels03epl,herr04pss,kohlbrecher05,michels05epl,michels05apl,
michels06prb,michels2010epjb,michels2012prb2}, amorphous alloys \cite{bracchi04,garcia04,cald05,mergia08}, and of elemental nanocrystalline bulk ferromagnets \cite{michels00a,michels00b,michels02a,michels02c,michels03prl,weissm04a,loeff05,michels08rop,
michels08epl,elmas09,michels2011jpcm,dobrichprb2012}, the process of dynamic nuclear polarization \cite{michels06a}, imaging of the flux-line lattice in superconductors \cite{forgan06,forgan2011}, precipitates in steels \cite{bischof07}, nanocrystalline rare-earth metals with random paramagnetic susceptibility \cite{weissm08}, fractal magnetic domain structures in NdFeB permanent magnets \cite{kreyssig09}, spin structures of ferrofluids, nanoparticles, and nanowires \cite{gazeau02,thomson05,ijiri05,albi06,grigorieva07,napolski07,bonini07,oku09,gri10,
krycka2010,disch2012}, magnetostriction in FeGa alloys \cite{laver2010}, electric-field-induced magnetization in multiferroics \cite{uehland2010}, magnetization reversal in magnetic recording media \cite{lister2010} and exchange-bias materials \cite{dufour2011}, and chiral and skyrmion-like structures in single crystals \cite{pappas09,gri09,pflei2009}.

Necessary prerequisite for the quantitative analysis of elastic magnetic SANS data is the knowledge of the Fourier components of the static magnetization vector field $\mathbf{M}(\mathbf{r})$ of the sample under study. The theory of micromagnetics \cite{brown,aharonibook,kronfahn03} provides the proper framework for the computation of $\mathbf{M}(\mathbf{r})$. However, the solution of Brown's equations of micromagnetics amounts to the solution of a set of nonlinear partial differential equations with complex boundary conditions, a task which cannot be done analytically for most practically relevant problems. Therefore, closed-form expressions for the ensuing so-called spin-misalignment scattering cross section are limited to the approach-to-saturation regime \cite{michels08rop,kron63}, in which the micromagnetic equations can be linearized.

In this review article we summarize our recent work, in which we have used \textsl{numerical} micromagnetics for the computation of the magnetic SANS cross section of two-phase magnetic nanocomposites. The use of numerical techniques allows us to solve the underlying equations rigorously, without resorting to the high-field approximation (saturation regime). This approach provides insights into the fundamentals of magnetic SANS. The micromagnetic simulations are adapted to the microstructure of a two-phase nanocomposite from the NANOPERM family of alloys \cite{suzuki06}.

The paper is organized as follows: in Sec.~2 we provide the details of our micromagnetic methodology, Sec.~3 discusses the magnetic SANS cross sections for the two most commonly used scattering geometries, Sec.~4 presents and discusses the results of the micromagnetic simulations for the magnetic SANS cross section, and in Sec.~5 we summarize the main findings of this study.

\section{Micromagnetic background}

Micromagnetism is a mesoscopic phenomenological theory designed to compute the equilibrium magnetization state of an arbitrarily shaped ferromagnetic body, when the applied field, the geometry of the ferromagnet and all materials parameters are known \cite{brown,aharonibook,kronfahn03}. In order to find the equilibrium magnetization configuration $\mathbf{M}_{\mathrm{eq}}(\mathbf{r})$, the total magnetic free energy of a ferromagnet should be considered as a functional of its magnetization state, $E_{\mathrm{tot}} = E_{\mathrm{tot}}[\mathbf{M}(\mathbf{r})]$. The state which delivers a (local) minimum to this functional corresponds to the required equilibrium magnetization configuration, so that the problem amounts to the minimization of the total energy functional. In the most common case $E_{\mathrm{tot}}$ contains contributions from the energy due to an external field, exchange, anisotropy and magnetodipolar interaction energies. Due to the nonlocal nature of the magnetodipolar interaction, almost all practically interesting problems can not be treated analytically, so that numerical minimization of $E_{\mathrm{tot}}[\mathbf{M}(\mathbf{r})]$ should be carried out. In the contemporary research landscape, numerical micromagnetics is a large and still continuously expanding field. Recent reviews on the micromagnetic state of the art can be found in the handbook Ref.~\cite{kronparkinhandbook07}. In this article we briefly discuss only those methodical aspects of numerical micromagnetics which are important for simulations of nanocomposite materials.

First, we would like to emphasize that such materials are one of the most complicated objects from the point of view of numerical simulations. The main difficulty is that they consist of at least two phases, and the boundaries between these phases are complicated curved surfaces; a typical example is a hard-soft nanocomposite consisting of magnetically hard (i.e., having a large magnetocystalline anisotropy) crystal grains surrounded by a magnetically soft matrix. Such a system is very difficult to simulate for the following reasons. The majority of modern numerical micromagnetic methods can be subdivided into two classes, the so-called finite-difference and finite-element methods (FDM and FEM)  \cite{kronparkinhandbook07}. In FDM the system under study is discretized into a regular translationally invariant (usually rectangular) grid. Such a discretization allows, first, the evaluation of the exchange field by simple finite-difference formulas, which are the finite-difference approximations for the corresponding second-order differential operator acting on the magnetization field $\mathbf{M}(\mathbf{r})$ (see below). Second, the translational invariance of a FDM grid enables the usage of the fast Fourier transformation (FFT) technique for the computation of the long-range magnetodipolar interaction field and energy. For a system discretized into $N$ cells, the FFT technique reduces the operation count for this energy from $\sim N^2$ (for a direct summation) to $\sim N \, \log N$. However, a serious disadvantage of a regular grid is a pure approximation for arbitrarily curved surfaces and boundaries. This is an important drawback for simulations of magnetic nanocomposites, because the adequate representation of the interphase boundaries for the accurate evaluation of associated exchange and magnetodipolar interactions between different phases is crucially important.

The second group of numerical methods widely used in micromagnetics---finite-element methods (FEM)---employ the discretization of the system under study into arbitrarily shaped tetrahedrons. The flexibility of this discretization type allows one to represent curved boundaries (including those between magnetically hard inclusions and the soft magnetic matrix) with any desired accuracy. However, the price to pay for this flexibility is high. First, computation of the exchange field requires now complicated methods designed for the accurate representation of second-order differential operators on irregular lattices. Second (and most important), it is no longer possible to use FFT for the magnetodipolar field evaluation. For this reason, highly sophisticated methods for the computation of this field are used in FEM simulations. These methods, which are based on the decomposition of magnetic potentials inside the ferromagnet and in the outer space, and the subsequent solution of the corresponding Poisson equations for these potentials on irregular grids \cite{kronparkinhandbook07} require a high programming effort and result in \textsl{iterative} algorithms for the evaluation of the dipolar field for a given magnetization configuration (in contrast to the FFT technique).

Another important limitation of finite-element methods is that they can only be employed in simulations with \textsl{open} boundary conditions (OBC), so that \textsl{periodic} boundary conditions (or PBC, routinely applied in simulations of extended thin films and bulk materials in order to eliminate strong finite-size effects) can not be used. The impossibility to apply PBC is a serious disadvantage in simulations of SANS experiments on nanocomposites, whereby the scattering intensity is sensitive to magnetization fluctuations in the bulk. Artificial surface demagnetizing effects arising in simulations with OBC might be very significant in this case, due to a relatively small simulation volume affordable even for modern computers. In addition, the suppression of these effects is especially important for nanocomposites containing a soft magnetic phase.

Another undesirable feature of a tetrahedron mesh is that hard magnetic grains must also be discretized into tetrahedrons, although in many cases the magnetization within a single grain is nearly homogeneous. This leads to a significant increase of the total number of finite elements required, resulting in a corresponding increase of the computation time; we refer the reader to Ref.~\cite{fidler05} for the discussion of this problem.

Due to all the reasons explained above, numerical micromagnetic simulations of SANS experiments on nanocomposites are very rare \cite{ogrin06,herr08pss}. Corresponding full-scale simulations of SANS measurements on a two-phase system have been reported, up to our knowledge, only in Ref.~\cite{ogrin06}, where the magnetization configuration of a longitudinal magnetic recording media film was modeled. Based on the experimental characterization of this material, the authors of \cite{ogrin06} have built a two-phase model for this system, where each magnetic grain consisted of a hard magnetic grain core and an essentially paramagnetic grain shell, having a very high susceptibility. The OOMMF code employing the standard FDM has been used \cite{oommf}, so that a very fine discretization ($0.3 \times 0.3 \times 0.3 \, \mathrm{nm}^3$ cells) had to be applied in order to reproduce the spherical shape of grain cores with a required accuracy. For this reason, only a rather limited number of grains ($\sim 50$) could be simulated. In addition, the exchange interaction both between the grains and within the soft magnetic matrix (represented by the merging grain shells) was neglected. Still, using several adjustable parameters, a satisfactory agreement of the simulated SANS intensity profile with experimental data was achieved.

The brief overview of the methodical problems presented here clearly shows that both a qualitative improvement of the micromagnetic simulation methodology and extensive numerical studies devoted to SANS experiments are highly desirable.

\subsection{New micromagnetic algorithm: mesh generation and its regular representation}
\label{meshgen}

For the reasons explained in the previous section and in order to perform accurate and efficient simulations of two-phase nanocomposites, we need to generate a polyhedron mesh with the following properties: (a) it should allow to represent each hard magnetic crystallite as a single finite element (because the magnetization inside such a crystallite is essentially homogeneous), (b) the mesh should allow for an arbitrarily fine discretization of the soft magnetic matrix in-between the hard grains (to account for the possible large variations of the magnetization direction between the hard grains), and (c) the shape of the meshing polyhedrons should be as close as possible to spherical, in order to ensure a good quality of a spherical dipolar approximation for the calculation of the magnetodipolar interaction energy, even for the nearest neighboring mesh elements.

A mesh consisting of polyhedrons satisfying all these requirements can be generated using two kinds of methods. First, there exist various modifications of a purely geometrical algorithm designed to obtain a random close packing of hard spheres \cite{jodrey85}. In these algorithms, the initial distribution of sphere centers is completely random. Then, at each step the worst overlap between two spheres is eliminated by pushing these spheres apart along the line connecting their centers. This procedure usually introduces new overlaps; however, these overlaps are usually smaller and are eliminated during the next steps, so that on the average the packing quality improves (the largest overlap present in the system decreases). The algorithm is robust and produces a random close packing of nonoverlapping spheres with any desired accuracy (see Ref.~\cite{jodrey85} for further details). Unfortunately, the computation time for this method increases with the number of elements $N$ as $\sim N^2$, so that the maximal number of spheres which can be positioned within a reasonable computation time is $N \sim 10^4$.

Therefore, in order to generate a mesh with a much larger number of finite elements ($N > 10^5$), we have developed a ``physical'' method, where we model a system of spheres interacting via a short-range repulsive potential:
\begin{equation}
\label{repulsivepotential}
U_i = \sum_{j = 1}^N A_{\mathrm{pot}} \, \exp \left\lbrace - \frac{d_{ij} - (r_i + r_j)}{r_{\mathrm{dec}}} \right\rbrace .
\end{equation}
Here, the constant $A_{\mathrm{pot}}$ determines the value of our potential when the distance $d_{ij}$ between the centers of interacting spheres is equal to the sum of their radii $r_i$ and $r_j$ (to ensure small overlaps in the final configuration, it should be $A_{\mathrm{pot}} \gg 1$; in a typical case $A_{\mathrm{pot}} = 10$). The parameter $r_{\mathrm{dec}}$ defines the decay radius of the potential. Again, at the beginning of iterations, sphere centers are positioned randomly. Then, we move the spheres according to the purely dissipative (i.e., neglecting the inertial term) equation of motion resulting from the forces obtained from the potential Eq.~(\ref{repulsivepotential}). The time step for the integration of this equation is adjusted to ensure decrease of the total system energy after each step. Due to the repulsive nature of the potential Eq.~(\ref{repulsivepotential}), this procedure leads also to the decrease of overlaps of the spheres. To achieve the desired result, we move the spheres until their maximum overlap does not exceed some prescribed small value (we have found that for our purposes the remaining overlap $(r_i + r_j)/d_{ij} > 0.95$ is good enough). The algorithm may be refined further to increase its efficiency; in particular, one might decrease the decay radius of the potential $r_{\mathrm{dec}}$, thus making the potential ``harder'', when the overlapping between spheres decreases during the sphere motion. We also note that due to the random spatial arrangement of spheres obtained in this way, we avoid possible artifacts caused by the regular placement of finite elements.

After the spheres have been positioned using one of the two algorithms described above, their centers are used as location points of magnetic dipoles. To compute the magnitudes $\mu_i$ of these dipoles, we have to determine the volume of each corresponding mesh element, which is in fact a polyhedron. This determination is made via a regular grid representation procedure that should satisfy the following requirements. First, we should conserve the total sample volume. Second, the interface between neighboring mesh elements should be flat as far as possible (apart from geometrical reasons). The last requirement is also supported by electron microscopy images of various polycrystalline magnets (e.g., \cite{gutfleisch2006,gutfleisch2010}).

In order to satisfy both these requirements, we used the following method: the sample is divided into cubical cells which side is much smaller (usually about four times smaller) than the size of a finite element (polyhedron) of our disordered mesh used to discretize the soft phase. For every cubical cell $(j,k,l)$, we calculate the distances $\Delta s^i_{j,k,l}$ between the center of this cell and the centers of neighboring polyhedrons (labeled by $i$). The function
\begin{equation}
\label{minreggrid}
\mathrm{min}_{\{i\}} \left[ (\Delta s^i_{j, k, l})^2 - R_i^2 \right]
\end{equation}
indicates to which polyhedron (with radius $R_i$) we attribute the current $(j,k,l)$ cube. The sum of cube volumes ascribed to the given polyhedron is taken as its volume. As a result of this procedure, the distribution of mesh-element volumes for both magnetic phases demonstrates a nearly Gaussian behavior. To obtain the magnitude of the dipolar moment assigned to each polyhedron, its volume is multiplied by the saturation magnetization of the material inside which the polyhedron is located (we remind that nanocomposites consist of materials with different magnetizations).

This method also allows for a very efficient calculation of the Fourier components of the magnetization (see Eqs.~(\ref{sigmasansperp}) and (\ref{sigmasanspara}) below) for a disordered system, using FFT on the already composed regular grid.

\begin{figure}
\centering
\resizebox{1.0\columnwidth}{!}{\includegraphics{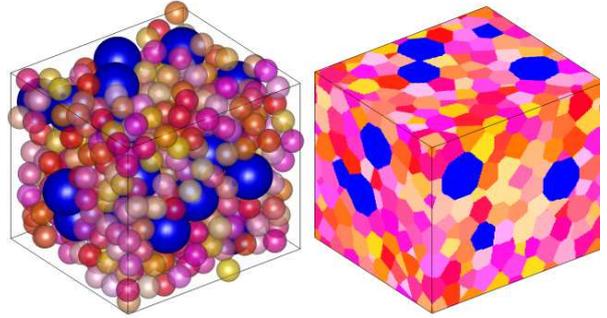}}
\caption{Schematical representation of the mesh-generation method: spheres in the left image indicate the distribution of magnetic dipoles (blue---hard magnetic phase, yellow-orange-red---soft magnetic phase). The corresponding regular grid representation on the right is used for the mesh-element volume determination (see Sec.~\ref{meshgen} for details). Note that the actual ``sample'', which is used for the micromagnetic simulations, is featured in Fig.~\ref{fig3}.}
\label{fig1}
\end{figure}

Summarizing, the whole algorithm can be viewed as a method to discretize a sample into polyhedrons having nearly spherical shape (see Fig.~\ref{fig1}). This is due to the fact that polyhedrons ``inherit'' the spatial structure obtained by the positioning of closely packed spheres. The fact that the shape of the volume which is occupied by each magnetic moment is nearly spherical allows us to use the spherical dipolar approximation (equivalent to the point dipole approximation) for the evaluation of the magnetodipolar interaction between the moments.

Finally, we point out that both algorithms allow for the usage of polyhedrons with different sizes, if we need different meshing on different system locations.

\subsection{New micromagnetic algorithm: energy contributions}

In our micromagnetic simulations we take into account all four standard contributions to the total magnetic free energy listed above: energy in the external magnetic field, energy of the magnetocrystalline anisotropy, exchange stiffness and magnetodipolar interaction energies.

\subsubsection{External field and magnetocrystalline anisotropy energies}

The system energy due to the presence of an external magnetic field and the energy of the magnetocrystalline anisotropy (which can be uniaxial and/or cubic) are calculated in our model in the standard way, namely
\begin{equation}
\label{externalenergy}
E_{\mathrm{ext}} = - \sum_{i = 1}^N \mbox{\boldmath$\mu$}_i \, \mathbf{H} ,
\end{equation}
\begin{equation}
\label{unianisotropyenergy}
E_{\mathrm{an}}^{\mathrm{un}} = - \sum_{i = 1}^N K_i^{\mathrm{un}} \, V_i \, \left( \mathbf{m}_i \, \mathbf{n}_i \right)^2 ,
\end{equation}
\begin{eqnarray}
\label{cubanisotropyenergy}
E_{\mathrm{an}}^{\mathrm{cub}} = \sum_{i = 1}^N K_i^{\mathrm{cub}} \, V_i \, \left( m^2_{i,x'} m^2_{i,y'} + m^2_{i,y'} m^2_{i,z'} + m^2_{i,x'} m^2_{i,z'} \right) ,
\end{eqnarray}
where $\mathbf{H}$ is the external field, $\mbox{\boldmath$\mu$}_i = \mbox{\boldmath$\mu$}(\mathbf{r}_i)$ and $V_i$ are the magnetic moment and the volume of the $i$th finite element (polyhedron), and $\mathbf{m}_i$ denotes the unit magnetization vector. Both the anisotropy constants $K_i$ and the directions of the anisotropy axes $\mathbf{n}_i$ can be site-dependent, as required for a polycrystalline nanocomposite material. For the cubic anisotropy case, the symbols $m_{i,x'}$ etc.\ represent the components of unit magnetization vectors in the local coordinate system that is attached to the cubic anisotropy axes.

\subsubsection{Exchange energy}

The evaluation of this energy contribution in our model requires a much more sophisiticated approach than in the standard FDM, because the continuous integral version of this energy contains the magnetization gradients,
\begin{equation}
\label{intexenergy}
E_{\mathrm{exch}} = \int_V A(\mathbf{r}) \, \left[ (\nabla m_x)^2 + (\nabla m_y)^2 + (\nabla m_z)^2 \right] \, dV ,
\end{equation}
where $A$ denotes the exchange-stiffness parameter and $V$ is the sample volume. Finding an approximation to Eq.~(\ref{intexenergy}) for a disordered system is a highly nontrivial task.

We remind that for a regular cubic grid with a cell size $a$ (and cell volume $\Delta V = a^3$), it can be shown rigorously (see the detailed proof in Ref.~\cite{berkov05handbook}) that the integral in Eq.~(\ref{intexenergy}) can be approximated as the sum
\begin{equation}
\label{regexenergy}
E_{\mathrm{exch}} = - \frac{1}{2} \sum_{i = 1}^N \sum_{j \subset \mathrm{n.n.}(i)} \frac{2A_{ij} \, \Delta V}{a^2} \, \left( \mathbf{m}_i \, \mathbf{m}_j \right).
\end{equation}
Here, $A_{ij}$ denotes the exchange-stiffness constant between cells $i$ and $j$, and the notation $j \subset \mathrm{n.n.}(i)$ means that the inner summation is performed over the nearest neighbors of the $i$th cell only. We note in passing that this Heisenberg-like expression is valid only when the angles between neighboring moments are not too large; as shown in Ref.~\cite{berkov05handbook}, neglecting this condition can lead to completely unphysical results.

For a disordered system of finite elements having different volumes, different distances between the element centers, and different numbers of nearest neighbors for each element, the expression Eq.~(\ref{regexenergy}) obviously can not be used. The most straightforward way to compute the exchange interaction in such a system would be to employ a rigorous numerical approximation suitable for the evaluation of the integral Eq.~(\ref{intexenergy}), where the integrand values (magnetization vectors $\mathbf{m}$) are given at arbitrarily placed spatial points. Derivation of such integration formulas amounts to the approximation of the magnetization projections $m_{x,y,z}(\mathbf{r})$ using some kind of a polynomial interpolation of these functions between the points where their values are defined (in our case, polyhedron centers). The integrand in Eq.~(\ref{intexenergy}) includes first spatial derivatives of the magnetization field, so that corresponding polynomials for the energy evaluation should deliver a continuous (and even better smooth) first derivative of $\mathbf{m}(\mathbf{r})$. For the effective field evaluation, where the continuous expression involves second-order derivatives, the polynomial interpolation is even more demanding. In addition, we should keep in mind that the condition $|\mathbf{m}(\mathbf{r})| = 1$ must be fulfilled everywhere, so that the interpolation of the magnetization angles, rather than that of Cartesian magnetization components should be used. All these features would result in a highly complicated algorithm for the exchange energy and field evaluation, which might be, in addition, subject to serious stability problems to usage of the angle interpolation.

For these reasons, we have decided to develop an algorithm for the exchange-energy evaluation based on the summation of the nearest neighbors contributions [similar to the expression (Eq.~\ref{regexenergy})] and at the same time take into account the differences between a regular grid and a disordered system mentioned above. To achieve this goal, we modify the expression Eq.~(\ref{regexenergy}) in the following way. First, from the derivation of the expression Eq.~(\ref{regexenergy}) presented in \cite{berkov05handbook}, it is clear that for the regular mesh consisting of cubic cells, the volume $\Delta V$ in the numerator of this expression is actually not the volume of the cell, but the volume enclosed between the centers of the cell $i$ and the neighboring cell $j$ (for a cubic lattice or a lattice consisting of rectangular prisms, this volume is obviously equal to the cell volume, because it includes two halves of identical cells). Therefore, for a disordered system of arbitrary finite elements, $\Delta V$ should be replaced by $\overline V_{ij} = (V_i + V_j)/2$ , where $V_i$ and $V_j$ are the volumes of the $i$th and the $j$th finite elements.

The second adjustment of Eq.~(\ref{regexenergy}) to a finite element system is the replacement of the distance $a$ between the cell centers in a regular lattice by the distance $\Delta r_{ij}$ between the centers of cells $i$ and $j$.

The third and most complicated correction is due to the different number of nearest neighbors in a regular lattice and in a disordered system of finite elements. In a regular Cartesian lattice, each cell has exactly $N_{\mathrm{nn}} = 6$ nearest neighbors, and the angles between the lines connecting the cell center with the centers of its neighbors in $x$, $y$ and $z$ directions are always $90^{\circ}$. For this reason, the overlapping of volumes enclosed between the centers of neighboring cells in, e.g., $x$ and $y$ directions is the same for all cells, what was taken into account by derivation of Eq.~(\ref{regexenergy}) (it is important to note, that this overlapping should not be confused with the overlapping of spheres mentioned in the discussion of algorithms used to contruct our disordered mesh in Sec.~\ref{meshgen}). In contrast to this nice feature of a regular Cartesian lattice, in a disordered system of finite elements the number of nearest neighbors for different finite elements may be different, and the overlapping of volumes enclosed between the centers of a given cell and its different neighbors may also vary. For example, for the element with more than six nearest neighbors, the volumes enclosed between its center and centers of its neighbors would overlap more than for a cubic lattice. For such an element, the exchange-stiffness energy evaluated using the sum Eq.~(\ref{regexenergy}) would be overestimated due to this excessive overlapping, even when the two corrections explained above would be taken into account.

The simplest method to solve this problem is the introduction of the correction factor $6/n_{\mathrm{av}}$, where $n_{\mathrm{av}}$ is the \textsl{average} number of nearest neighbors for the particular random realization of our disordered finite-element system. This correction would compensate on average the effect of the incorrect count of overlapping regions explained above. The accuracy of this simple correction method can be hardly estimated in advance, but both simple tests performed in \cite{erokhin2011ieee} and additional much more complicated tests discussed in Sec.~\ref{minprocedure} below show that the accuracy provided by this correction method is surprisingly good.

Summarizing, for magnetic moments belonging to the same phase we propose the following expression for the exchange-stiffness energy:
\begin{equation}
\label{exonephaseenergy}
E_{\rm exch} = - \frac{1}{2} \sum_{i = 1}^N \sum_{j \subset \mathrm{n.n.}(i)} \frac{2A_{ij} \, \overline V_{ij}}{\Delta r_{ij}^2} \left( \mathbf{m}_i \, \mathbf{m}_j \right) ,
\end{equation}
where $\overline V_{ij} = (V_i + V_j)/2$, $\Delta r_{ij}$ is the distance between the centers of the $i$th and the $j$th finite elements with volumes $V_i$ and $V_j$, and $A_{ij}$ is the exchange constant.

The last point to be discussed is the choice of nearest neighbors, which should be used in the inner summation in Eq.~(\ref{exonephaseenergy}). The choice whether two elements should be considered as nearest neighbors is not unambiguous in disordered systems. We have adopted the following convention: two magnetic moments are considered as nearest neighbors, if the distance between the centers of corresponding polyhedrons is not larger than $d_{\mathrm{max}} = 1.4 \, (r_i + r_j)$. The cut-off factor $f_{\mathrm{cut}} = 1.4$ is chosen so that for the overwhelming majority of finite elements those two of them which have a common face are treated as nearest neighbors.

To evaluate the exchange-interaction energy between two finite elements (polyhedrons) belonging to \textsl{different} phases (hard and soft), we use the formula:
\begin{equation}
\label{extwophaseenergy}
E_{\mathrm{exch}} = - \frac{1}{2} \sum_{i = 1}^N \sum_{j \subset \mathrm{n.n.}(i)} \frac{2A_{ij} \, V_{\mathrm{sp}}/2}{(\Delta r_{ij} - R_{\mathrm{hp}})^2}(\mathbf{m}_i \, \mathbf{m}_j).
\end{equation}
Here, $V_{\mathrm{sp}}$ is the volume of a soft-phase element and $R_{\mathrm{hp}}$ is the radius of the sphere corresponding to the hard phase polyhedron. This modified expression Eq.~(\ref{extwophaseenergy}) accounts for the fact that in this case the magnetization rotation occurs almost entirely within the polyhedron corresponding to the soft phase.

\subsubsection{Magnetodipolar interaction}

The energy of the long-range magnetodipolar interaction between magnetic moments and the corresponding contribution to the total effective field are computed using the point-dipole approximation as
\begin{equation}
\label{dipenergy}
E_{\mathrm{dip}} = - \frac{1}{2} \sum_{i = 1}^N \mbox{\boldmath$\mu$}_i \sum_{j \ne i} \frac{3 \mathbf{e}_{ij} \, (\mathbf{e}_{ij} \, \mbox{\boldmath$\mu$}_j) - \mbox{\boldmath$\mu$}_j} {\Delta r^{3}_{ij}} ,
\end{equation}
i.e., magnetic moments of finite elements are treated as point dipoles located at the polyhedron centers. This approximation is equivalent to the approximation of spherical dipoles, i.e., it would be exact for spherical finite elements. Hence, for our discretized system, this approximation introduces some computational errors, because our finite elements are polyhedrons. However, these errors are small, because the shape of these polyhedrons is close to spherical (see Fig.~\ref{fig3}), due to the special algorithm employed for the generation of our mesh, as explained in Sec.~\ref{meshgen}.

The summation in Eq.~(\ref{dipenergy}) is performed by the so called particle-mesh Ewald method. Didactically very instructive and detailed introduction into Ewald methods can be found in Ref.~\cite{gibbon02}. The specific implementation of the lattice-based Ewald method for the magnetodipolar interaction for regular and disordered systems of magnetic particles is described in our papers \cite{berkov98,gorn07}. Here, we briefly repeat the basic issues of this algorithm to make our paper self-containing.

First we remind that the Ewald method \cite{hockneybook} was initially invented for evaluating conditionally converging lattice sums for the Coulomb interaction in ionic crystals. At present, it is a standard method to calculate any long-range interaction---including Coulomb sums, gravitation energy, dipole interaction, elastic forces in dislocation networks etc.---in systems with periodic boundary conditions (PBC). In such systems direct summation over all field sources is impossible simply due to their infinite number. Hence, we must use a Fourier expansion over the reciprocal lattice vectors $\mathbf{k}$ which correspond to the infinitely repeated simulation volume. For the \textsl{point} sources of the \textsl{long-range} field, the corresponding Fourier components decay relatively slowly with increasing magnitude of the wave vector $k$ in reciprocal space. In numerical simulations we always have to our disposal only a finite number of such wave vectors, so that the Fourier spectrum of our long-range interaction should be cut off at the maximal finite value $k_{\mathrm{max}}$. As mentioned above, the Fourier harmonics decay slowly, so that at $k_{\mathrm{max}}$ they are by no means small. For this reason, the spectrum cut-off due to the elimination of all Fourier components with $k > k_{\mathrm{max}}$ is sharp, thus, leading to large artificial oscillations of the interaction potential after its inverse transformation to the real space.

As with nearly all Ewald methods, the version described below for dipolar systems solves the problem by adding and subtracting a Gaussian dipole at each location of a point dipole $\mbox{\boldmath$\mu$}_i$ in the initial system. Using the definition of the gradient of the $\delta$ function, it is easy to show that this operation corresponds to the addition and subtraction of a charge distribution (with a width $\sigma$)
\begin{equation}
\label{charge_distrib}
\rho _i^{\mathrm{G}}(\mathbf{r}) =  - \frac{(\mathbf{r} - \mathbf{r}_i) \, \mbox{\boldmath$\mu$}_i}{(2\pi )^{3/2} \, \sigma^5} \, \exp \left(- \frac{ (\mathbf{r} - \mathbf{r}_i)^2 }{2\sigma ^2} \right) .
\end{equation}

Then the magnetodipolar field ${\mathbf{H}}^{\mathrm{dip}}  = \mathbf{H}_{\mathrm{A}}^{\mathrm{dip}}  + \mathbf{H}_{\mathrm{B}}^{\mathrm{dip}}$ is evaluated as the sum of two contributions from subsystems A and B. The first subsystem consists of Gaussian dipoles Eq.~(\ref{charge_distrib}) and the second one is composed of the original point dipoles minus these Gaussian dipoles,
\begin{equation}
\label{gauss_dipoles}
\rho_{\mathrm{B}}(\mathbf{r}) = - \sum_{i = 1}^N \left[ \mbox{\boldmath$\mu$}_i \, \nabla \delta(\mathbf{r} - \mathbf{r}_i) - \rho_i^{\mathrm{G}}(\mathbf{r}) \right] ,
\end{equation}
where the first terms in the sum on the right represent the charge density of a point dipole located at $\mathbf{r}_i$. The field created by a composite object in square brackets of Eq.~(\ref{gauss_dipoles}) is \cite{berkov98}
\begin{eqnarray}
\label{Halpha}
H_{\mathrm{B}, i}^\alpha(\mathbf{r} - \mathbf{r}_i) = \left[ \frac{3(\alpha - \alpha_i) \, (\mbox{\boldmath$\mu$}_i \, \Delta \mathbf{r}_i)}{\Delta r_i^5} - \frac{\mu_i^\alpha}{\Delta r_i^3} \right] \, f_{\mathrm{G}}(\Delta r_i) + \nonumber \\
\sqrt{\frac{2}{\pi}} \frac{(\alpha - \alpha_i) \, (\mbox{\boldmath$\mu$}_i \, \Delta \mathbf{r}_i)}{\Delta r_i^5} \, \exp \left[- \frac{\Delta r_i^2}{2\sigma^2} \right] , \nonumber \\
\end{eqnarray}
where $\alpha = x, y, z$.

It is important to note that the function $f_{\mathrm{G}}(r)$ decays with distance as $\exp(-r^2$),
\begin{equation}
\label{fGr}
f_{\mathrm{G}}(r) = 1 - \mathrm{erf}\left( \frac{r}{\sigma \sqrt 2} \right) + \frac{\sqrt{2} r}{\pi a} \, \exp \left[- \frac{r^2}{2\sigma^2} \right] .
\end{equation}

The goal of this decomposition of the original system of \textsl{point} dipoles is the following. The field Eq.~(\ref{Halpha}) from the second subsystem B is a \textsl{short-range} one, because each point dipole is screened by a Gaussian dipole with the same total moment, but with the opposite sign. The computation of such a short-range contribution takes $\sim N$ operations for a system of $N$ particles. The first subsystem A consists of dipoles having a smooth Gaussian charge distribution, so that its Fourier components decay rapidly with increasing $k$. This fast decay allows a painless cut-off of the Fourier spectrum at large wave vectors, so that the contribution from the first subsystem Eq.~(\ref{charge_distrib}) can be safely calculated using Fourier expansion. More detailed explanations concerning this procedure can be found in Ref.~\cite{berkov98}.

Already the above most straightforward implementation of the Ewald method allows for a reliable evaluation of the dipolar field in systems with PBC. However, for disordered systems, this method has the same prohibitively high operation count $\sim N^2$, as a direct summation for systems with OBC. The reason is that  particle positions in disordered systems do not form a regular lattice, so that the Fourier transformation for the calculation of the long-range contribution $\mathbf{H}_{\mathrm{A}}^{\mathrm{dip}}$ can not be done via the \textsl{fast} Fourier transformation technique: exponential factors $\exp(\mathrm{i} \mathbf{k} \mathbf{r}_i)$ should be computed for all wave vectors $\mathbf{k}$ and all particle positions $\mathbf{r}_i$ separately, leading to the operation count given above.

In order to decrease the computational costs, several lattice versions of the Ewald method have been developed (see the overview \cite{deserno98}). The general idea behind all these methods is to employ some mapping of the initial disordered system onto a regular lattice, in order to enable the application of the FFT. Using this general paradigm, we have implemented the following algorithm:

$\mathbf(i)$ First, we map our disordered system of point magnetic dipoles $\mbox{\boldmath$\mu$}_i = \mbox{\boldmath$\mu$}(\mathbf{r}_i)$ onto a system of dipoles located at lattice points $\mathbf{r_p}$ ($\mathbf{p}$ is the 3D index) using some weighting function  $w_{3d}(\mathbf{r})$,
\begin{eqnarray}
\label{w3d}
\mbox{\boldmath$\tilde{\mu}$}(\mathbf{r_p}) = \sum_{i = 1}^N \mbox{\boldmath$\mu$}(\mathbf{r}_i) \, w_{3d}(|\mathbf{r}_i - \mathbf{r}_p|) = \nonumber \\
\sum_{i = 1}^{M_{\mathrm{nb}}} \mbox{\boldmath$\mu$}_i \, w(|x_i - x_p|) \, w(|y_i - y_p|) \, w(|z_i - z_p|).
\end{eqnarray}
We emphasize that the whole method makes only sense if the mapping function $w$ is strongly localized, so that the sum over all $N$ dipoles in Eq.~(\ref{w3d}) is actually restricted to a few nearest neighbors $M_{\mathrm{nb}}$ of the lattice node $\mathbf{p}$.

$\mathbf(ii)$ Next, we add and subtract to each point \textsl{lattice} dipole two Gaussian dipoles Eq.~(\ref{gauss_dipoles}), as in the straightforward Ewald method described above.

$\mathbf(iii)$ Further, we compute the dipolar field of this lattice system as described above, i.e., as the sum of the long-range contribution from smooth Gaussian dipoles positioned \textsl{on the lattice} and the short-range contribution Eq.~(\ref{Halpha}) from the composite objects ``point dipole $-$ Gaussian dipole'', also placed on the lattice.

$\mathbf(iv)$ Finally, the field obtained in this way on the \textsl{lattice} points $\mathbf{r_p}$ is mapped back onto the initial dipole locations $\mathbf{r}_i$ using the same functions $w$ as in Eq.~(\ref{w3d}).

As mentioned above, the major advantage of this lattice Ewald version is the possibility to use FFT for computing the long-range part of the total magnetodipolar field. In addition, we can also accelerate the evalutaion of the short-range contribution. Namely, we note that $(i)$ the contribution Eq.~(\ref{Halpha}) depends only on the difference $\Delta \mathbf{r}$ between the source and target coordinates and $(ii)$ both source and target points are located on the lattice. Hence, this short-range contribution also is a discrete convolution and as such can be also computed by the FFT technique. Using this nice feature, we can increase the number of nearest neighbor shells used by the evaluation of the short-range interaction part without additional time cost, making the corresponding truncation error arbitrarily small. Keeping in mind that the evaluation of the long-range field part via the FFT technique for the \textsl{lattice} system is exact, we conclude that the only source of computational errors in our algorithm is the mapping of the initial disordered system onto a lattice, which can be easily controlled and reduced by choosing the suitable mapping scheme \cite{deserno98}. We have found that already the conventional first-order mapping used together with a lattice having a cell size equal to $R_{\mathrm{sp}}/2$ (here $R_{\mathrm{sp}}$ is the sphere radius used to generate the mesh for the soft-phase discretization) ensures by the evaluation of $\mathbf{H}^{\mathrm{dip}}$ a relative error smaller than $0.01$, which is good enough for our purposes.

\subsection{New micromagnetic algorithm: minimization procedure and numerical tests}
\label{minprocedure}

For the minimization of the total magnetic energy, obtained as the sum of all contributions described above, we use the simplified version of a gradient method employing the dissipation part of the Landau-Lifshitz equation of motion for magnetic moments \cite{kronparkinhandbook07,landau35}. This means that we update the magnetization configuration at each step as
\begin{equation}
\label{minproc}
\mathbf{m}_i^{\mathrm{new}} = \mathbf{m}_i^{\mathrm{old}} - \Delta t \left[ \mathbf{m}_i^{\mathrm{old}} \times \left[ \mathbf{m}_i^{\mathrm{old}} \times \mathbf{h}_i^{\mathrm{eff}} \right] \right],
\end{equation} 			
where $\mathbf{m}_i$ denotes the unit magnetization vector $\mathbf{m}_i = \mathbf{M}_i/M_S$ and $\mathbf{h}_i^{\mathrm{eff}}$ is the reduced effective field, evaluated in a standard way as the negative energy derivative over the magnetic moment projections \cite{kronparkinhandbook07}.

Since we are looking for the energy minimum, the time step in Eq.~(\ref{minproc}) is chosen and adapted using the monitoring of this energy. If the total energy decreases after the iteration step performed according to Eq.~(\ref{minproc}), we accept this step. If the energy increases, we restore the previous magnetization state, halve the time step ($\Delta t \rightarrow \Delta t/2$) and repeat the iteration. During the minimization procedure we may also increase the time step to avoid an unnecessary slow minimization: the time step is doubled, if the last few steps (typically $5 - 10$ steps) were successful. For the termination of the minimization procedure, we use the local torque criterion: we stop the iteration process, if the maximal torque acting on magnetic moments is smaller than some prescribed value, i.e., $\mathrm{max}_i\{|\mathbf{m}_i \times \mathbf{h}_i^{\mathrm{eff}}|\} < \varepsilon$. As is well known, this condition is more appropriate than the alternative criterion of a sufficiently small energy difference between the two subsequent steps. In all tested cases the value $\varepsilon = 10^{-3}$ was small enough to ensure the minimization convergence.

The new methodology explained in detail above was tested on two simple examples in Ref.~\cite{erokhin2011ieee}. We remind that we have first reproduced---using our disordered mesh---with a high accuracy the analytically known magnetization profile of a standard 3D Bloch wall. Second, for a trial 3D magnetic configuration defined via simple trigonometric functions of coordinates (we have used these functions to ensure a sufficiently slow spatial variation of the system magnetization direction), we have obtained a very good agreement between the total energy and partial energy contributions found by our new method and the FDM micromagnetic package MicroMagus \cite{micromagus}.

\begin{figure*}
\centering
\resizebox{1.50\columnwidth}{!}{\includegraphics{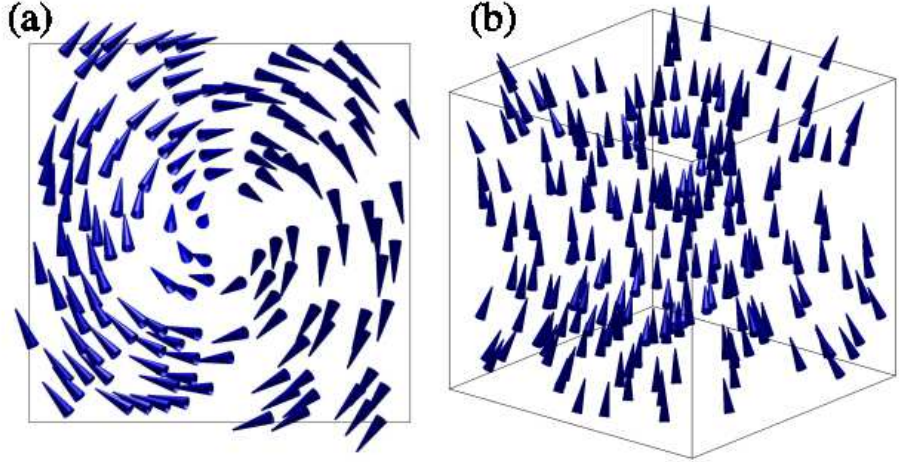}}
\caption{(a) Vortex (2D cross section) and (b) flower (3D arrow plot) magnetization states obtained by our new methodology explained in Sec.~\ref{minprocedure} above. Data taken from Ref.~\cite{erokhin2012prb}.}
\label{fig2}
\end{figure*}

Here, we would like to present two additional much more complicated tests, where we compute the equilibrium magnetization configurations of a cubic magnetic particle obtained using our new method and compare these configurations with the results obtained for the same system by the MicroMagus package.

The particle size was chosen to be $40 \times 40 \times 40 \, \mathrm{nm}^3$, and the magnetic materials parameters were set to $M_S = 800 \, \mathrm{kA/m}$, $A = 1.0 \times 10^{-11} \, \mathrm{J/m}$, and $K = 5.0 \times 10^4 \, \mathrm{J/m^3}$ (uniaxial anisotropy). For the simulations using our new method, the particle was discretized into $N = 9000$ polyhedrons with a typical size of $d = 2 \, \mathrm{nm}$. For the MicroMagus simulations, a cubic cell with a side length of $2.5 \, \mathrm{nm}$ was used.

For the test problems, we have chosen two well known magnetization states typical for ferromagnetic particles of this size \cite{rave98}: the vortex state and the so-called flower state. To obtain the vortex state, we have started the minimization procedure from the state that is topologically equivalent to the vortex, the so-called closed Landau domain configuration. The flower state could be obtained by starting the energy minimization simply from the homogeneous configuration with the magnetization directed along a cube side.

\begin{table}
\caption{Comparison of energies and reduced magnetizations for the vortex and flower magnetization states computed by the new method and by the standard finite difference simulations (MicroMagus software). Data taken from Ref.~\cite{erokhin2012prb}.}
%\centering
\footnotesize
\begin{tabular}{lll}
\hline
Vortex energies ($\times 10^{-18} \, \mathrm{J}$) & New method & MicroMagus \\
\hline
$E_{\mathrm{tot}}$  & 8.225 & 8.270 \\
$E_{\mathrm{an}}$   & 1.361 & 1.385 \\
$E_{\mathrm{exch}}$ & 4.409 & 4.562 \\
$E_{\mathrm{dip}}$  & 2.455 & 2.324 \\
$M/M_S$             & 0.400 & 0.406 \\
\hline \\
\hline
Flower energies ($\times 10^{-18} \, \mathrm{J}$) & New method & MicroMagus \\
\hline
$E_{\mathrm{tot}}$  & 7.813 & 7.843 \\
$E_{\mathrm{an}}$   & 0.137 & 0.127 \\
$E_{\mathrm{exch}}$ & 0.434 & 0.441 \\
$E_{\mathrm{dip}}$  & 7.242 & 7.275 \\
$M/M_S$             & 0.972 & 0.974 \\
\hline
\end{tabular}
\label{test_table}
\end{table}

Table~\ref{test_table} lists the total energies, partial energy contributions and the reduced magnetization values for the equilibrium magnetization states shown in Fig.~\ref{fig2} obtained by the new method and by the standard FDM simulations (MicroMagus package). Almost all energy contributions obtained by the two methods agree very well. The only significant relative difference can be found for the anisotropy energy of the flower state; however, this significant \textsl{relative} difference ($\Delta E/E$) arises simply due to a very low value of this energy. All in one, the agreement between the new and the established methodologies for all cases where the standard methods are applicable is fully satisfactory.

\section{Magnetic SANS cross section of unpolarized neutrons}

In our micromagnetic simulations of elastic magnetic SANS we have focussed on the two most commonly employed scattering geometries where the wavevector $\mathbf{k}_0$ of the incident neutron beam is either perpendicular [case~(i)] or parallel [case~(ii)] to the external magnetic field $\mathbf{H}$, which is applied along the $z$-direction of a Cartesian coordinate system. ($\mathbf{e}_x$, $\mathbf{e}_y$ and $\mathbf{e}_z$ represent the unit vectors along the Cartesian axes.) Furthermore, since the focus of the present study is on magnetic spin-misalignment scattering, we have ignored the nuclear SANS contribution. Note, however, that for polycrystalline texture-free magnetic nanocomposites the nuclear SANS signal is virtually independent of the applied magnetic field and isotropic, and its magnitude is generally small compared to the here relevant spin-misalignment scattering \cite{michels08rop}. Furthermore, the restriction to unpolarized neutrons entails the neglect of scattering contributions from helical spin arrangments, which are of relevance, e.g., in FeCoSi and MnSi single crystals \cite{gri09,pflei2009}.

\begin{figure}[t!]
\centering
\resizebox{1.0\columnwidth}{!}{\includegraphics{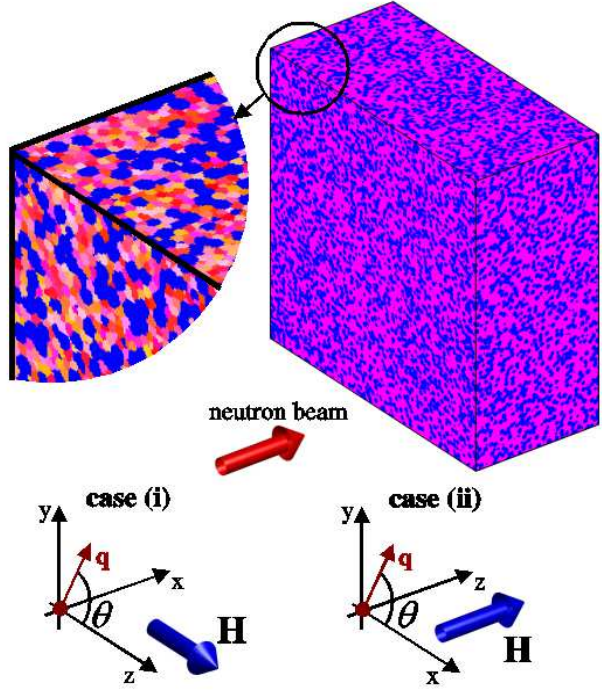}}
\caption{Sketch of the two scattering geometries and of the microscopic structure of the nanocomposite sample (simulation volume: $V = 250 \times 600 \times 600 \, \mathrm{nm}^3$). Blue polyhedrons---Fe particles; yellow-orange-red polyhedrons---matrix phase. In the micromagnetic simulations the sizes of the Fe particles are distributed according to a Gaussian function centered at about $10 \, \mathrm{nm}$ and the particle volume fraction equals $40 \, \%$, as in the experimental study Ref.~\cite{michels06prb}.}
\label{fig3}
\end{figure}

A sketch of the above two scattering geometries along with a schematic drawing of the microscopic structure of the nanocomposite sample can be seen in Fig.~\ref{fig3}.

\subsection{Case~(i): $\mathbf{k}_0 \perp \mathbf{H} \parallel \mathbf{e}_z$}

For $\mathbf{k}_0 \parallel \mathbf{e}_x$, the elastic magnetic SANS cross section $d \Sigma_M / d \Omega$ at momentum-transfer vector $\mathbf{q}$ reads \cite{michels08rop}
\begin{eqnarray}
\label{sigmasansperp}
\frac{d \Sigma_M}{d \Omega}(\mathbf{q}) = \frac{8 \pi^3}{V} b_{\mathrm{H}}^2 \left( |\widetilde{M}_x|^2 + |\widetilde{M}_y|^2 \cos^2\theta \right. \nonumber \\ \left. + |\widetilde{M}_z|^2 \sin^2\theta - (\widetilde{M}_y \widetilde{M}_z^{\ast} + \widetilde{M}_y^{\ast} \widetilde{M}_z) \sin\theta \cos\theta \right),
\end{eqnarray}
where $V = 250 \times 600 \times 600 \, \mathrm{nm}^3$ is the scattering volume, $b_{\mathrm{H}} = 2.9 \times 10^{8} \, \mathrm{A}^{-1} \mathrm{m}^{-1}$, $c^{*}$ is a quantity complex-conjugated to $c$, and $\widetilde{M}_{(x,y,z)}(\mathbf{q})$ are the Fourier transforms of the magnetization components $M_{(x,y,z)}(\mathbf{r})$. Note that in the small-angle limit and for this particular geometry the scattering vector $\mathbf{q}$ can be expressed as $\mathbf{q} \cong q \, (0, \sin\theta, \cos\theta)$, where $\theta$ denotes the angle between $\mathbf{q}$ and $\mathbf{H}$ (Fig.~\ref{fig3}).

\subsection{Case~(ii): $\mathbf{k}_0 \parallel \mathbf{H} \parallel \mathbf{e}_z$}

For this geometry, one finds
\begin{eqnarray}
\label{sigmasanspara}
\frac{d \Sigma_M}{d \Omega}(\mathbf{q}) = \frac{8 \pi^3}{V} b_{\mathrm{H}}^2 \left( |\widetilde{M}_x|^2 \sin^2\theta + |\widetilde{M}_y|^2 \cos^2\theta \right. \nonumber \\ \left. + |\widetilde{M}_z|^2 - (\widetilde{M}_x \widetilde{M}_y^{\ast} + \widetilde{M}_x^{\ast} \widetilde{M}_y) \sin\theta \cos\theta \right),
\end{eqnarray}
where $\mathbf{q} \cong q \, (\cos\theta, \sin\theta, 0)$ and $\theta$ is measured relative to $\mathbf{e}_x$ (Fig.~\ref{fig3}).

For the micromagnetic simulations of magnetic SANS from two-phase nanocomposites, we used (unless otherwise stated) the following materials parameters for hard (``h'') and soft (``s'') phases, which are characteristic for the Fe-based nanocrystalline alloy NANOPERM \cite{suzuki06}: magnetizations $M_{\mathrm{h}} = 1750 \, \mathrm{kA/m}$ and $M_{\mathrm{s}} = 550 \, \mathrm{kA/m}$, anisotropy constants $K_{\mathrm{h}} = 4.6 \times 10^4 \, \mathrm{J/m^3}$ and $K_{\mathrm{s}} = 1.0 \times 10^2 \, \mathrm{J/m^3}$. As a value for the exchange-stiffness constant we used $A = 0.2 \times 10^{-11} \, \mathrm{J/m}$ for interactions both within the soft phase and between the hard and soft phases.

\section{Results and discussion}

\begin{figure*}[ht!]
\centering
\resizebox{1.25\columnwidth}{!}{\includegraphics{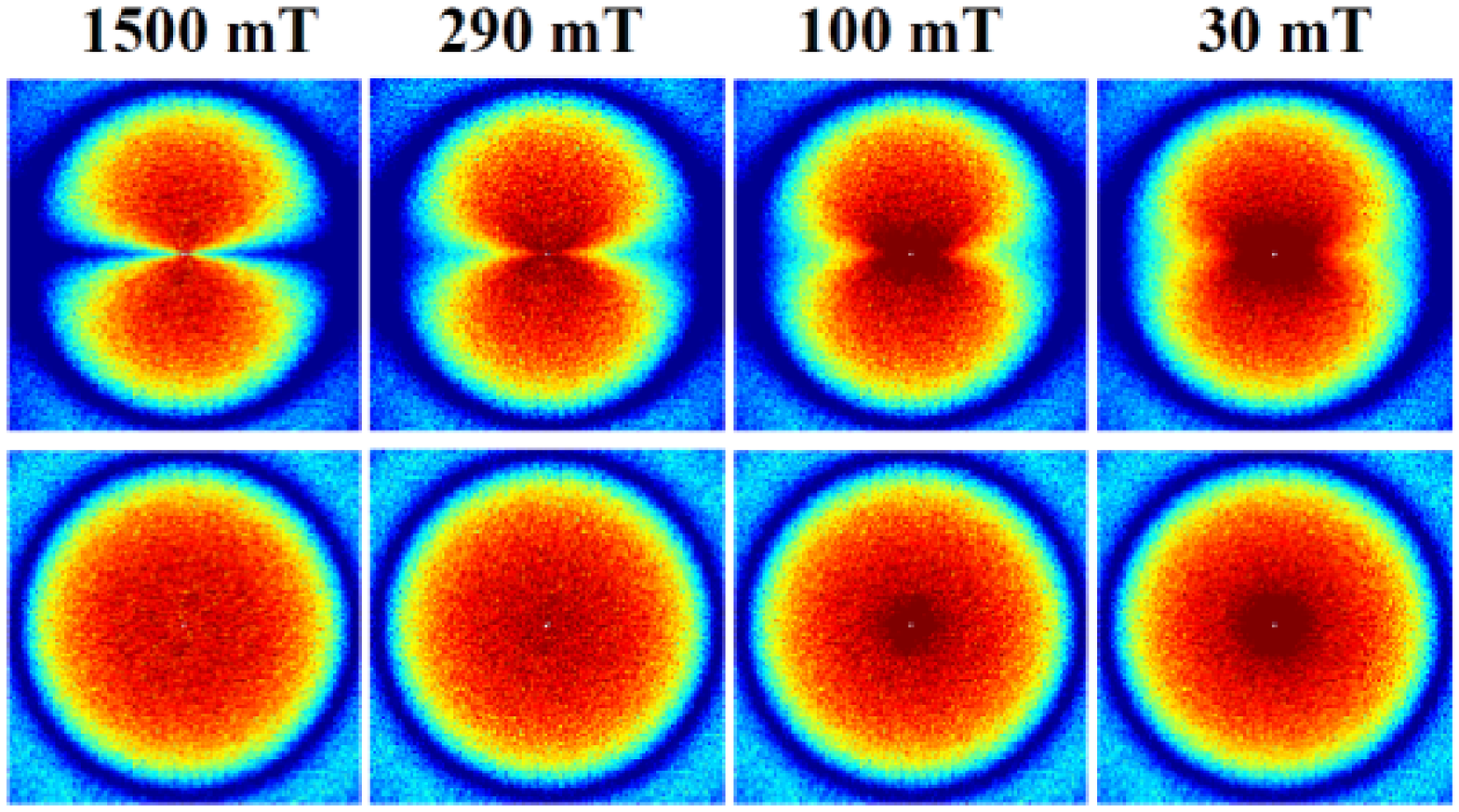}}
\caption{Applied-field dependence of the total magnetic SANS cross section $d \Sigma_M / d \Omega$ for $\mathbf{k}_0 \perp \mathbf{H}$ (Eq.~(\ref{sigmasansperp}), upper row) and for $\mathbf{k}_0 \parallel \mathbf{H}$ (Eq.~(\ref{sigmasanspara}), lower row). The external magnetic field $\mathbf{H} \parallel \mathbf{e}_z$ is applied horizontally in the plane of the detector for $\mathbf{k}_0 \perp \mathbf{H}$ ($q_x = 0$, upper row) and normal to the detector plane for $\mathbf{k}_0 \parallel \mathbf{H}$ ($q_z = 0$, lower row). Materials parameters of NANO\-PERM were used (see text). Pixels in the corners of the images have $q \cong 1.2 \, \mathrm{nm}^{-1}$. Logarithmic color scale is used.}
\label{fig4}
\vspace{0.50cm}
%\end{figure}
%\begin{figure}[h!]
\centering
\resizebox{1.25\columnwidth}{!}{\includegraphics{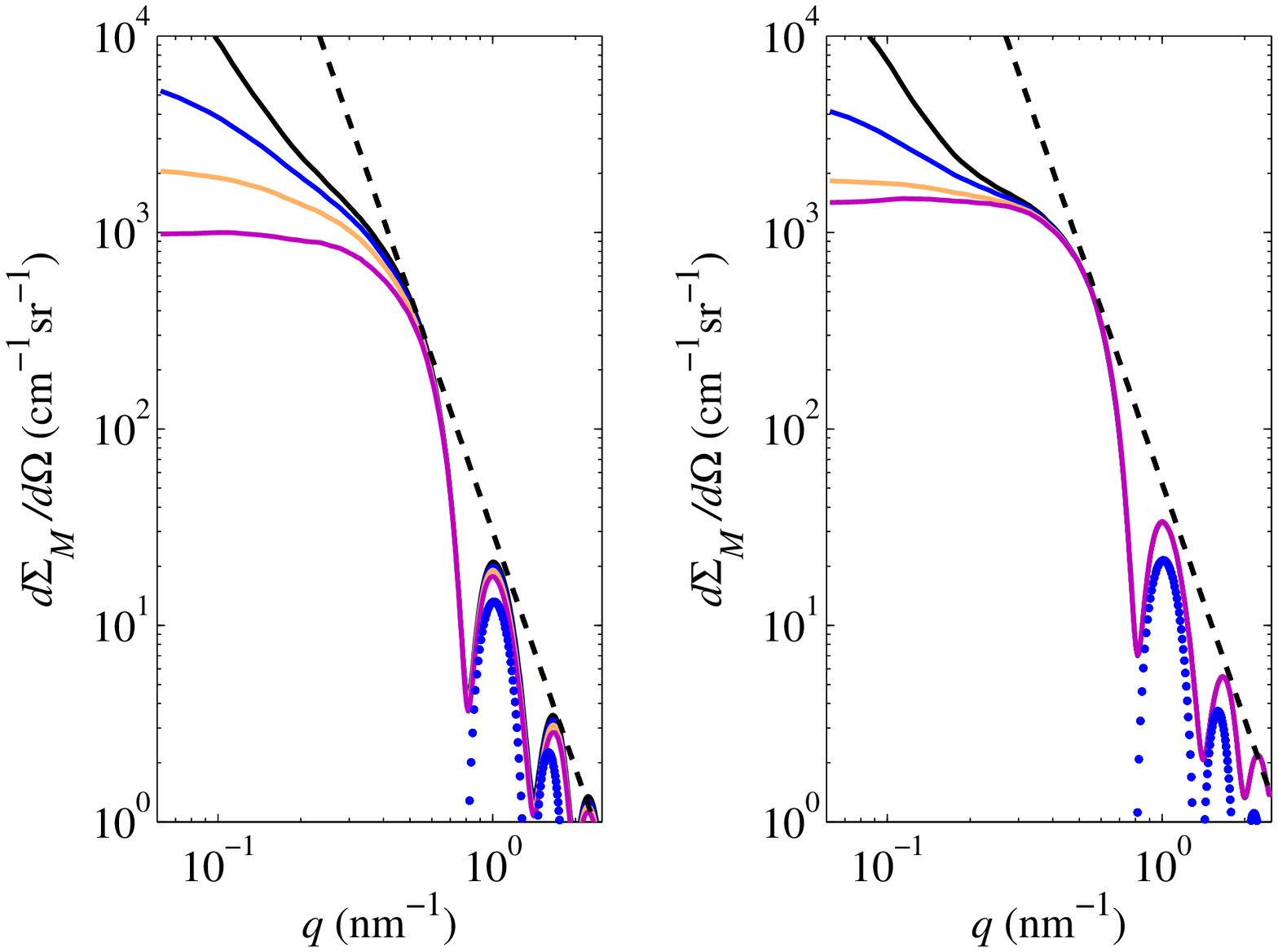}}
\caption{Solid lines: Radially-averaged $d \Sigma_M / d \Omega$ as a function of scattering vector $q$ for $\mathbf{k}_0 \perp \mathbf{H}$ (left image) and for $\mathbf{k}_0 \parallel \mathbf{H}$ (right image) (data have been smoothed). Field values (in mT) from top to bottom, respectively: 30, 100, 290, 1500. Dashed lines in both images: $d \Sigma_M / d \Omega \propto q^{-4}$. Solid circles in both images represent part of the (squared) form factor of a sphere with a radius of $R = 5.7 \, \mathrm{nm}$.}
\label{fig5}
\end{figure*}

The applied-field dependence of the total magnetic SANS cross sections $d \Sigma_M / d \Omega$ [computed, respectively, by means of Eqs.~(\ref{sigmasansperp}) and (\ref{sigmasanspara})] is displayed in Fig.~\ref{fig4} for both scattering geometries, i.e., for the situations when the wavevector $\mathbf{k}_0$ of the incoming neutron beam is perpendicular [case~(i)] or parallel [case~(ii)] to the applied magnetic field $\mathbf{H}$, which for both cases is assumed to be parallel to $\mathbf{e}_z$. The corresponding radially-averaged data can be seen in Fig.~\ref{fig5}.

While $d \Sigma_M / d \Omega$ for $\mathbf{k}_0 \parallel \mathbf{H}$ is isotropic (i.e., $\theta$ independent) over the whole field and momentum-transfer range, it is highly anisotropic for $\mathbf{k}_0 \perp \mathbf{H}$ (Fig.~\ref{fig4}). At a saturating applied magnetic field of $\mu_0 H = 1.5 \, \mathrm{T}$, where the normalized ``sample'' magnetization is (for both geometries) larger than $99.9 \, \%$, the anisotropy of $d \Sigma_M / d \Omega$ [case~(i)] is clearly of the $\sin^2\theta$-type, i.e., elongated normal to $\mathbf{H}$; this is because magnetic scattering due to transversal spin misalignment is small close to saturation and the dominating scattering contrast arises from nanoscale jumps of the longitudinal magnetization at phase boundaries. On decreasing the field, the transversal magnetization components increase in magnitude as long-range spin misalignment develops at the smallest $q$. The SANS pattern in case~(i) essentially remains of the $\sin^2\theta$-type at lower fields, although a more complicated anisotropy builds up at small $q$. As can be seen in Fig.~\ref{fig5}, $d \Sigma_M / d \Omega$ at small $q$ increases by more than one order of magnitude as the field is decreased from $1.5 \, \mathrm{T}$ to $30 \, \mathrm{mT}$. Asymptotically, at large $q$, the power-law dependence of $d \Sigma_M / d \Omega$ can be described by $d \Sigma_M / d \Omega \propto q^{-4}$. In agreement with the nature of the underlying microstructure ($\sim 10 \, \mathrm{nm}$-sized single-domain Fe particles in a nearly saturated matrix), one can describe the oscillations of $d \Sigma_M / d \Omega$ at large $q$ and $H$ by the form factor of a sphere with a radius of $R = 5.7 \, \mathrm{nm}$ (solid circles in Fig.~\ref{fig5}). We remind that the shape of the particles in our micromagnetic algorithm is not strictly spherical.

\begin{figure*}[ht!]
\centering
\resizebox{2.0\columnwidth}{!}{\includegraphics{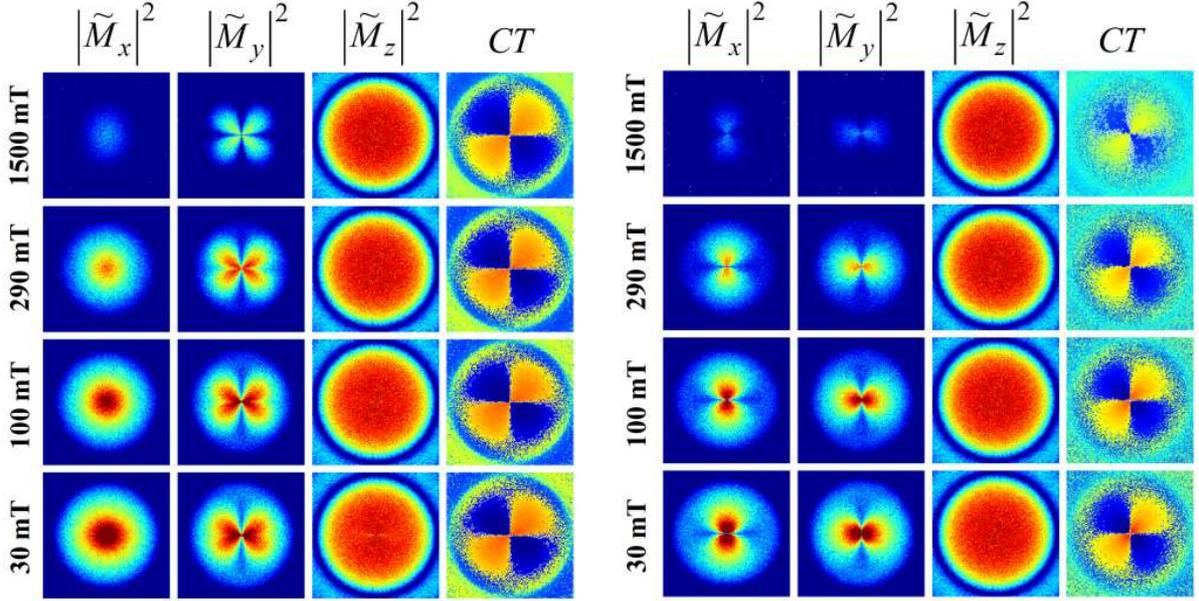}}
\caption{Results of the micromagnetic simulations for the Fourier coefficients of the magnetization. The images represent projections of the respective functions into the plane of the 2D detector, i.e., $q_x = 0$ for $\mathbf{k}_0 \perp \mathbf{H}$ (left image) and $q_z = 0$ for $\mathbf{k}_0 \parallel \mathbf{H}$ (right image). From left column to right column, respectively: $|\widetilde{M}_x|^2$, $|\widetilde{M}_y|^2$, $|\widetilde{M}_z|^2$, and $CT = - (\widetilde{M}_y \widetilde{M}_z^{\ast} + \widetilde{M}_y^{\ast} \widetilde{M}_z)$ (left image) and $CT = - (\widetilde{M}_x \widetilde{M}_y^{\ast} + \widetilde{M}_x^{\ast} \widetilde{M}_y)$ (right image). In the first three columns from left, red color corresponds, respectively, to ``high intensity'' and blue color to ``low intensity''; in the fourth column, blue color corresponds to negative and orange-yellow color to positive values of the $CT$. Pixels in the corners of the images have $q \cong 1.2 \, \mathrm{nm}^{-1}$. Logarithmic color scale is used.}
\label{fig6}
\end{figure*}

Figure~\ref{fig6} shows the projections of the magnetization Fourier coefficients $|\widetilde{M}_x|^2$, $|\widetilde{M}_y|^2$, $|\widetilde{M}_z|^2$, and of the cross terms $CT = - (\widetilde{M}_y \widetilde{M}_z^{\ast} + \widetilde{M}_y^{\ast} \widetilde{M}_z)$ and $CT = - (\widetilde{M}_x \widetilde{M}_y^{\ast} + \widetilde{M}_x^{\ast} \widetilde{M}_y)$ into the plane of the 2D detector at the same external-field values as in Fig.~\ref{fig4}. It can be seen in Fig.~\ref{fig6} that in case~(i) both $|\widetilde{M}_x|^2$ and $|\widetilde{M}_z|^2$ are isotropic over the displayed ($q, H$) range, while the Fourier coefficient $|\widetilde{M}_y|^2$ reveals a pronounced angular anisotropy, with maxima that lie roughly along the diagonals of the detector (the so-called ``clover-leaf'' anisotropy, see Fig.~\ref{fig11} below). In case~(ii), $|\widetilde{M}_x|^2$ and $|\widetilde{M}_y|^2$ are both strongly anisotropic (with characteristic maxima in the plane perpendicular to $\mathbf{H}$), while $|\widetilde{M}_z|^2$ is isotropic. When [for case~(ii)] all Fourier coefficients are multiplied by the corresponding trigonometric functions and summed up [compare Eq.~(\ref{sigmasanspara})], the resulting $d \Sigma_M / d \Omega$ becomes isotropic (Fig.~\ref{fig4}, lower row).

\begin{figure}[htb]
\centering
\resizebox{0.6\columnwidth}{!}{\includegraphics{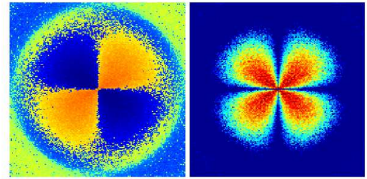}}
\caption{(left image) $CT = - (\widetilde{M}_y \widetilde{M}_z^{\ast} + \widetilde{M}_y^{\ast} \widetilde{M}_z)$ for $\mathbf{k}_0 \perp \mathbf{H}$ ($\mu_0 H = 0.29 \, \mathrm{T}$). Blue color corresponds to negative and orange-yellow color to positive values of the $CT$. (right image) $CT \sin\theta \cos\theta$. Red color corresponds to ``high intensity'' and blue color to ``low intensity''. All other settings are as in Fig.~\ref{fig6}.}
\label{fig7}
\end{figure}

The cross terms for both scattering geometries vary in sign between quadrants on the detector. The respective $CT$ is positive in the upper right quadrant of the detector ($0^{\circ} < \theta < 90^{\circ}$), negative in the upper left quadrant ($90^{\circ} < \theta < 180^{\circ}$), and so on. When both $CT$'s are multiplied by $\sin\theta \cos\theta$, the corresponding contribution to $d \Sigma_M / d \Omega$ becomes positive-definite for all angles $\theta$ (compare Fig.~\ref{fig7}). This obervation suggests that---contrary to the common assumption that the $CT$ averages to zero for statistically isotropic polycrystalline microstructures---the $CT$ appears to be of special relevance in nanocomposite magnets.

\begin{figure*}[t!]
\centering
\resizebox{1.20\columnwidth}{!}{\includegraphics{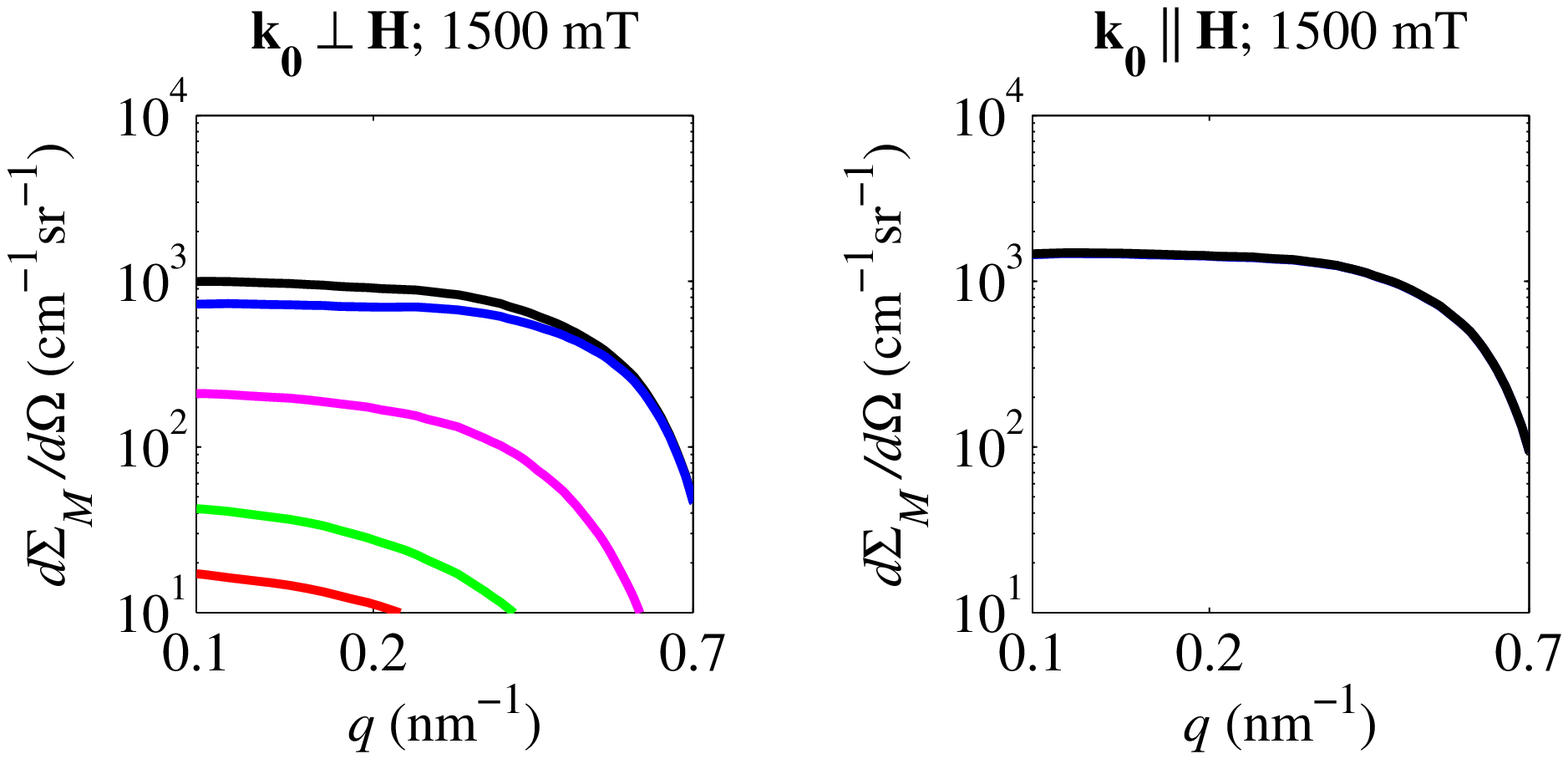}} \\ \vspace{0.25cm}
\resizebox{1.20\columnwidth}{!}{\includegraphics{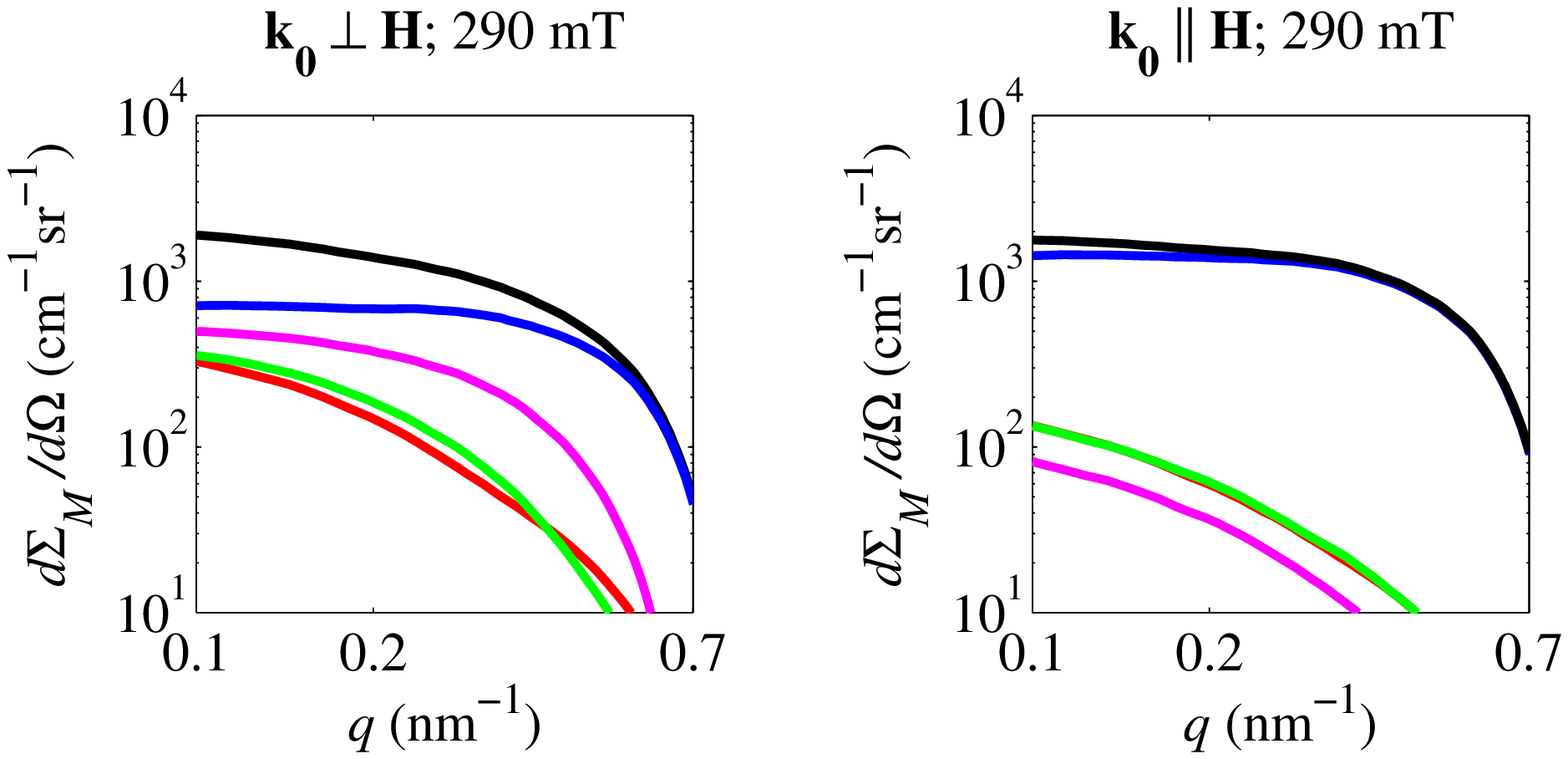}} \\ \vspace{0.25cm}
\resizebox{1.20\columnwidth}{!}{\includegraphics{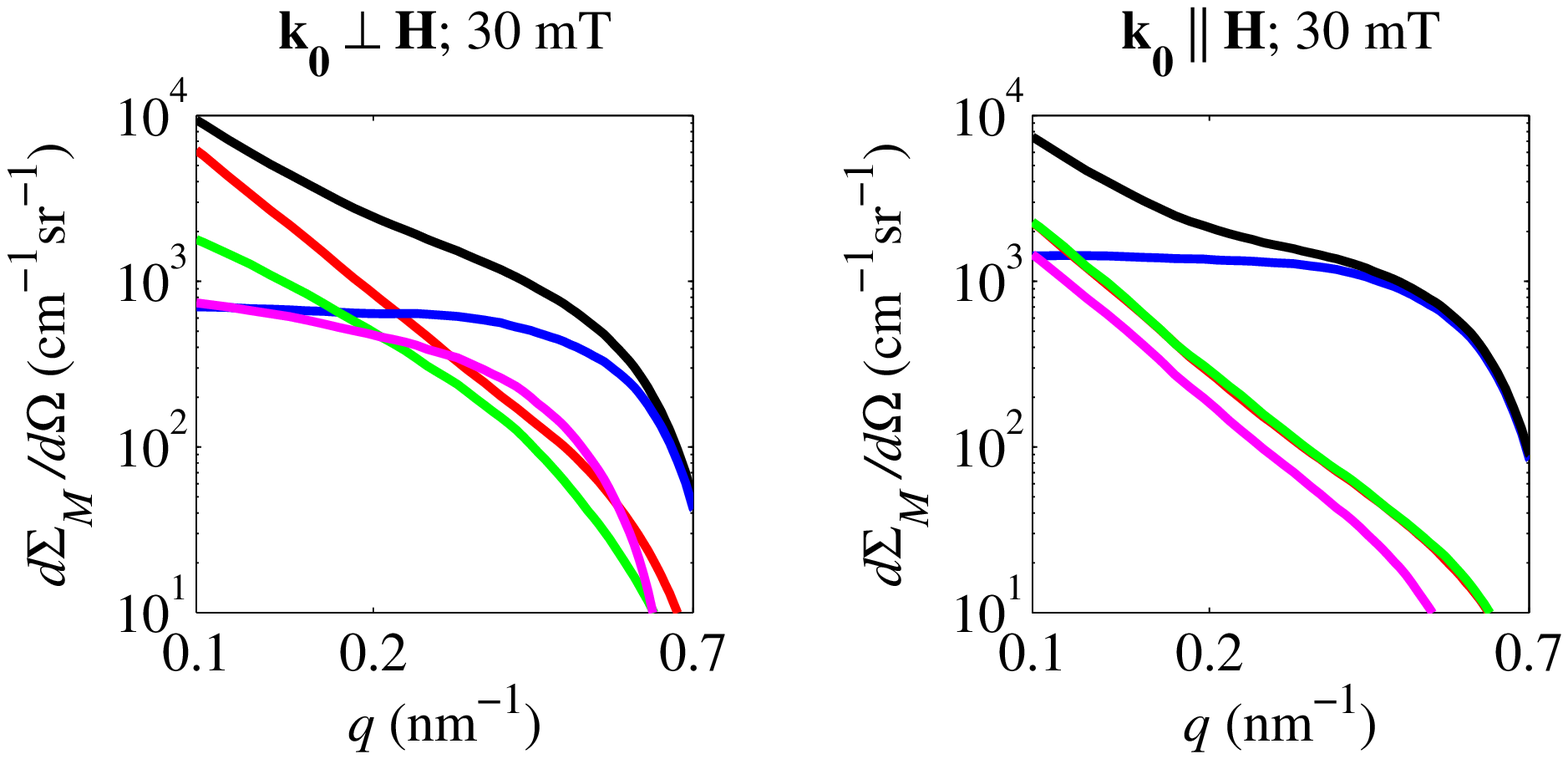}}
\caption{Radially-averaged total magnetic SANS cross sections $d \Sigma_M / d \Omega$ and radially-averaged individual contributions to $d \Sigma_M / d \Omega$ as a function of scattering vector $q$ and applied magnetic field $H$ for $\mathbf{k}_0 \perp \mathbf{H}$ and $\mathbf{k}_0 \parallel \mathbf{H}$ (see insets). Black lines: total $d \Sigma_M / d \Omega$; Blue lines: $|\widetilde{M}_z|^2$; Magenta lines: $CT$; Green lines: $|\widetilde{M}_y|^2$; Red lines: $|\widetilde{M}_x|^2$. In the above notation, the prefactor $\frac{8 \pi^3}{V} b_H^2$ and the numerical factors that result from the averaging procedure have been omitted for clarity. Note that different trigonometric functions may be involved in the averaging procedure, compare, e.g., $\frac{8 \pi^3}{V} b_H^2 |\widetilde{M}_x|^2$ for $\mathbf{k}_0 \perp \mathbf{H}$ [Eq.~(\ref{sigmasansperp})] and $\frac{8 \pi^3}{V} b_H^2 |\widetilde{M}_x|^2 \sin^2\theta$ for $\mathbf{k}_0 \parallel \mathbf{H}$ [Eq.~(\ref{sigmasanspara})].}
\label{fig8}
\end{figure*}
In Fig.~\ref{fig8} we show for both scattering geometries the radially-averaged total $d \Sigma_M / d \Omega$ along with the radially-averaged \textsl{individual} scattering contributions to $d \Sigma_M / d \Omega$, i.e., the radial average of terms $\frac{8 \pi^3}{V} b_H^2 |\widetilde{M}_x|^2$, $\frac{8 \pi^3}{V} b_H^2 |\widetilde{M}_y|^2 \cos^2\theta$, $\frac{8 \pi^3}{V} b_H^2 CT \sin\theta \sin\theta$, and so on [compare Eqs.~(\ref{sigmasansperp}) and (\ref{sigmasanspara})]. At saturation ($\mu_0 H = 1.5 \, \mathrm{T}$), both transversal scattering contributions, i.e., terms $\propto |\widetilde{M}_x(q)|^2$ and $\propto |\widetilde{M}_y(q)|^2$, are for both cases~(i) and (ii) small relative to the other terms and the main contribution to the total $d \Sigma_M / d \Omega$ originates from longitudinal magnetization fluctuations, i.e., from terms $\propto |\widetilde{M}_z(q)|^2$. For $\mathbf{k}_0 \parallel \mathbf{H}$, both transversal terms are so small that they are not visible within the displayed ``intensity'' range and $d \Sigma_M / d \Omega$ practically equals the $|\widetilde{M}_z(q)|^2$ scattering (both curves superimpose). Note that the $CT$ for case~(i) is the product of a transversal and the longitudinal magnetization Fourier coefficient, whereas for case~(ii) the $CT$ contains the two transversal components. This explains why the $CT$ for case~(ii) is much smaller than the $CT$ for case~(i) at fields close to saturation. On decreasing the field, the transversal Fourier coefficients and the $CT$'s become progressively more important, in particular at small $q$.

It is also important to note that the present simulations were carried out by assuming a quite large jump in the magnetization magnitude $\Delta M$ at the interphase between particles and matrix, $\mu_0 \Delta M = 1.5 \, \mathrm{T}$. Consequently, the ensuing $|\widetilde{M}_z(q)|^2$ scattering in both geometries and the $CT$ scattering in case~(i) are relatively large. For $\Delta M = 0$, the $CT$ at saturation for case~(i) becomes negligible, since $\widetilde{M}_z(\mathbf{q}) \propto \delta(\mathbf{q} = 0)$.

Figures~\ref{fig6} and \ref{fig8} embody the power of our approach: By employing numerical micromagnetics for the computation of magnetic SANS cross sections, it becomes possible to study the \textsl{individual} magnetization Fourier coefficients and their contribution to $d \Sigma_M / d \Omega$. This sheds light on the ongoing discussion regarding the explicit $\mathbf{q}$-dependence of $d \Sigma_M / d \Omega$ \cite{dufour2011}. In particular, the approach of combining micromagnetics and SANS complements neutron experiments, which generally provide only a weighted sum of Fourier coefficients [compare Eqs.~(\ref{sigmasansperp}) and (\ref{sigmasanspara})], a fact that often hampers the straightforward interpretation of recorded SANS data. While it is in principle possible to determine some Fourier coefficients, e.g., through the application of a saturating magnetic field or by exploiting the neutron-polarization degree of freedom via so-called SANSPOL or POLARIS methods (e.g., Refs.~\cite{michels2010epjb,krycka2010,albi2010}), it is difficult to unambiguously determine a particular scattering contribution without ``contamination'' by unwanted Fourier components. For instance, when the applied field is not large enough to completely saturate the sample, then the scattering of unpolarized neutrons along the field direction does not represent the pure nuclear SANS, but contains also the magnetic SANS due to the misaligned spins \cite{bischof07}.

\begin{figure}[t!]
\centering
\resizebox{1.0\columnwidth}{!}{\includegraphics{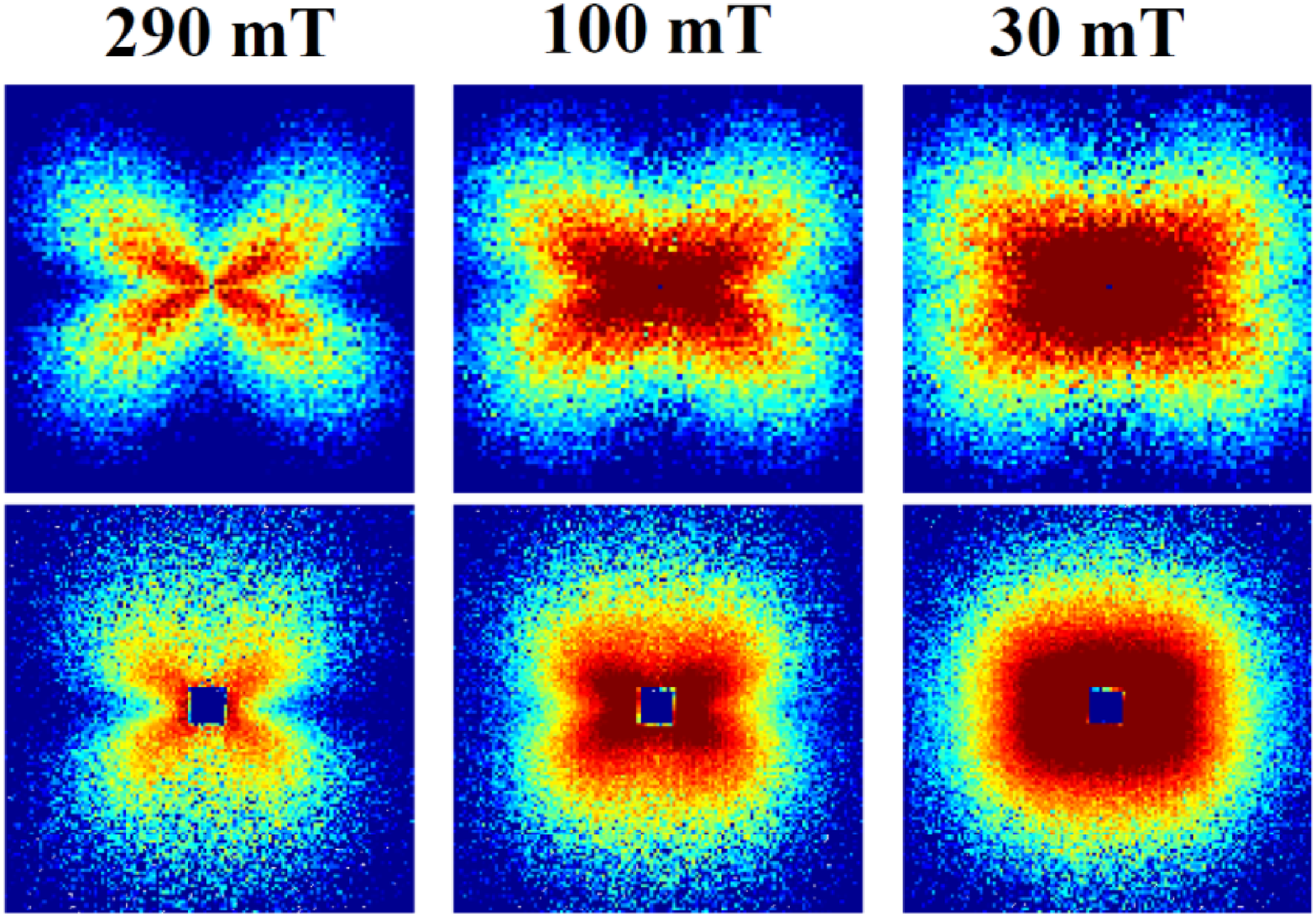}}
\caption{Comparison between simulation (upper row) and experimental data (lower row) for the difference cross section $\propto (|\widetilde{M}_x|^2 + |\widetilde{M}_y|^2 \cos^2\theta + CT \sin\theta \cos\theta)$ at different external fields as indicated ($\mathbf{k}_0 \perp \mathbf{H}$). Pixels in the corners of the images have $q \cong 0.64 \, \mathrm{nm}^{-1}$. Logarithmic color scale is used. Since the experimental data was not obtained in absolute units, we have multiplied it with a scaling factor for comparison with the simulated data. $\mathbf{H}$ is horizontal in the plane. Experimental data were taken from Ref.~\cite{michels06prb}.}
\label{fig9}
\vspace{0.50cm}
%\end{figure}
%\begin{figure}[h!]
\centering
\resizebox{1.0\columnwidth}{!}{\includegraphics{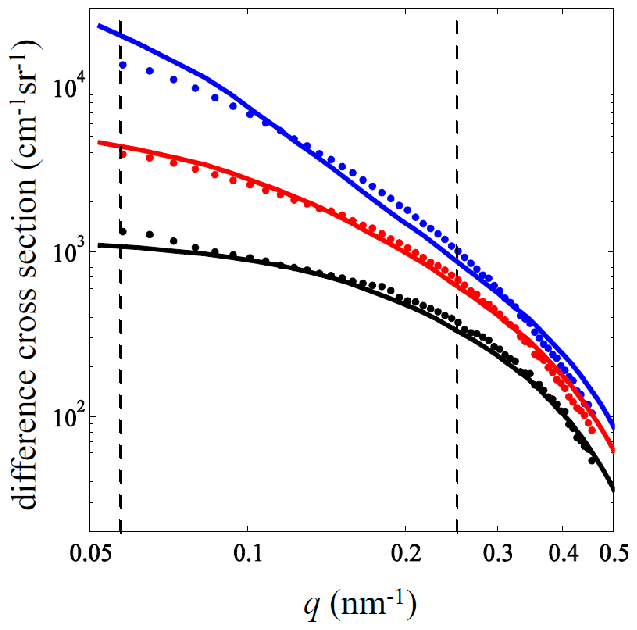}}
\caption{($\bullet$) Radially-averaged experimental difference cross sections as a function of momentum transfer $q$ and $H$ ($\mathbf{k}_0 \perp \mathbf{H}$). Field values (in mT) from top to bottom: 30, 100, 290. Solid lines: Results of the micromagnetic simulations (data have been smoothed). Vertical dashed lines indicate the region where the clover-leaf anisotropy is observed.  Experimental data were taken from Ref.~\cite{michels06prb} and multiplied by a scaling factor (compare Fig.~\ref{fig9}).}
\label{fig10}
\end{figure}

The finding [for case~(i)] that $|\widetilde{M}_x|^2$ and $|\widetilde{M}_z|^2$ are isotropic and that $|\widetilde{M}_y|^2 = |\widetilde{M}_y|^2(q, \theta)$ provides a straightforward explanation for the experimental observation of the clover-leaf anisotropy in the SANS data of the nanocrystalline two-phase alloy NANO\-PERM \cite{michels06prb}. Our simulation results for the difference cross section $\propto (|\widetilde{M}_x|^2 + |\widetilde{M}_y|^2 \cos^2\theta + CT \sin\theta \cos\theta)$ (see Figs.~\ref{fig9} and \ref{fig10}), where the scattering at saturation ($\mu_0 H = 1.5 \, \mathrm{T}$) has been subtracted, agree qualitatively well with the experimental data \cite{erokhin2011ieee,erokhin2012prb}. Clover-leaf-type anisotropies in $d \Sigma_M / d \Omega$ have also been reported for a number of other materials, including precipitates in steels \cite{bischof07}, nanocrystalline Gd \cite{michels08epl,dobrichprb2012}, and nanoporous Fe \cite{elmas09}.

The maxima in the difference cross section [for case~(i)] depend on $q$ and $H$, and may appear at angles $\theta$ significantly smaller than $45^{\circ}$. This becomes evident in Fig.~\ref{fig11}, where we show (for $\mathbf{k}_0 \perp \mathbf{H}$) polar plots of the simulated difference cross section at selected $q$ and $H$.
\begin{figure}[t!]
\centering
\resizebox{1.0\columnwidth}{!}{\includegraphics{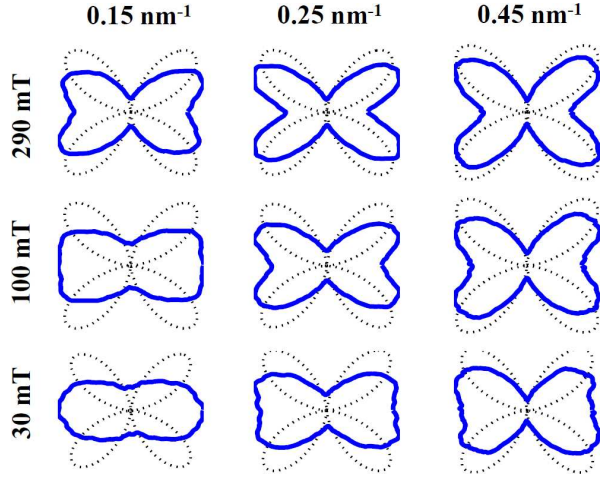}}
\caption{Polar plots of the simulated difference cross section $\propto (|\widetilde{M}_x|^2 + |\widetilde{M}_y|^2 \cos^2\theta + CT \sin\theta \cos\theta)$ at different combinations of momentum transfer $q$ and applied magnetic field $H$ (see insets) ($\mathbf{k}_0 \perp \mathbf{H}$). Data have been smoothed. Dotted lines ($\propto \sin^2 \theta \cos^2 \theta$) serve as guides to the eyes.}
\label{fig11}
\end{figure}

\begin{figure}[h!]
\centering
\resizebox{1.0\columnwidth}{!}{\includegraphics{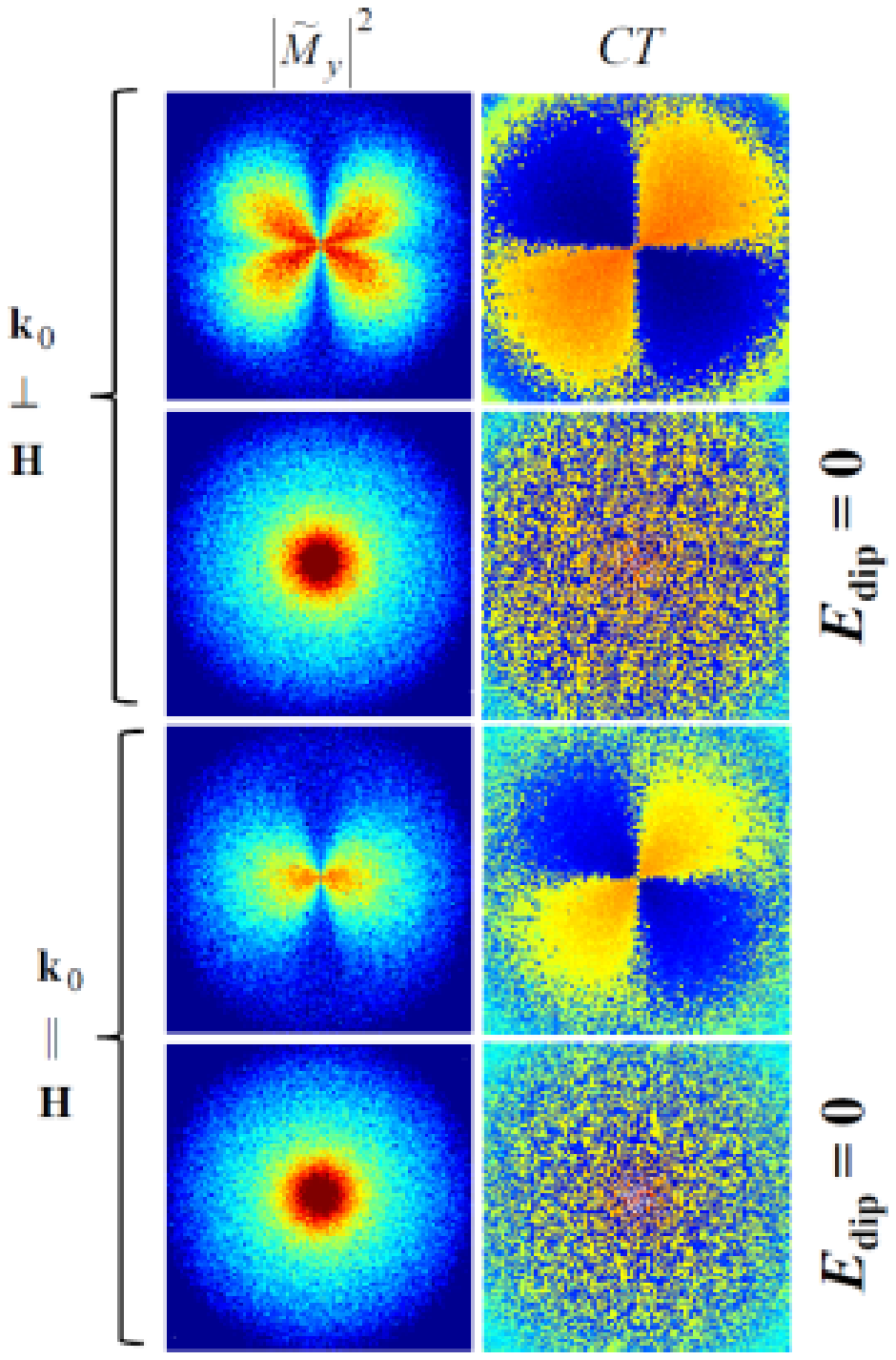}}
\caption{Influence of the dipolar interaction on the Fourier coefficients of the magnetization. $|\widetilde{M}_y|^2$ and both cross terms $CT = - (\widetilde{M}_y \widetilde{M}_z^{\ast} + \widetilde{M}_y^{\ast} \widetilde{M}_z)$ ($\mathbf{k}_0 \perp \mathbf{H}$) and $CT = - (\widetilde{M}_x \widetilde{M}_y^{\ast} + \widetilde{M}_x^{\ast} \widetilde{M}_y)$ ($\mathbf{k}_0 \parallel \mathbf{H}$) were computed from a real-space magnetic microstructure with a normalized magnetization of $99.0 \, \%$. Applied fields of $290 \, \mathrm{mT}$ (with dipolar interaction) and $7 \, \mathrm{mT}$ (without dipolar interaction $E_{\mathrm{dip}} = 0$) were required in order to achieve this magnetization value. The corresponding results for the Fourier coefficient $|\widetilde{M}_x|^2$ (for $\mathbf{k}_0 \parallel \mathbf{H}$) are analogous to the depicted results for $|\widetilde{M}_y|^2$. Pixels in the corners of the images have $q \cong 0.9 \, \mathrm{nm}^{-1}$. Logarithmic color scale is used.}
\label{fig12}
\end{figure}

The results of our previous work \cite{erokhin2012prb,michels2012prb1} strongly suggest that the magnetodipolar interaction plays a decisive role for the understanding of magnetic SANS of nanocomposites. In fact, it is this interaction which is responsible for the anisotropy, i.e., for the $\theta$-dependence of the magnetization Fourier coefficients and, hence, of $d \Sigma_M / d \Omega$. The impact of the dipolar interaction on $d \Sigma_M / d \Omega$ can be conveniently studied, since our micromagnetic algorithm allows one to ``switch on'' and ``off'' this energy term. Figure~\ref{fig12} shows results of micromagnetic simulations for $|\widetilde{M}_y|^2$ and for both $CT$'s obtained with and without the dipolar interaction. When the dipolar interaction is ignored in the micromagnetic computations, all Fourier coefficients are isotropic at all $q$ and $H$ investigated (data for $|\widetilde{M}_x|^2$ and $|\widetilde{M}_z|^2$ are not shown). This observation shows that for any realistic description of experimental magnetic SANS data this interaction has to be taken into account.

\begin{figure*}[ht]
\centering
\resizebox{1.50\columnwidth}{!}{\includegraphics{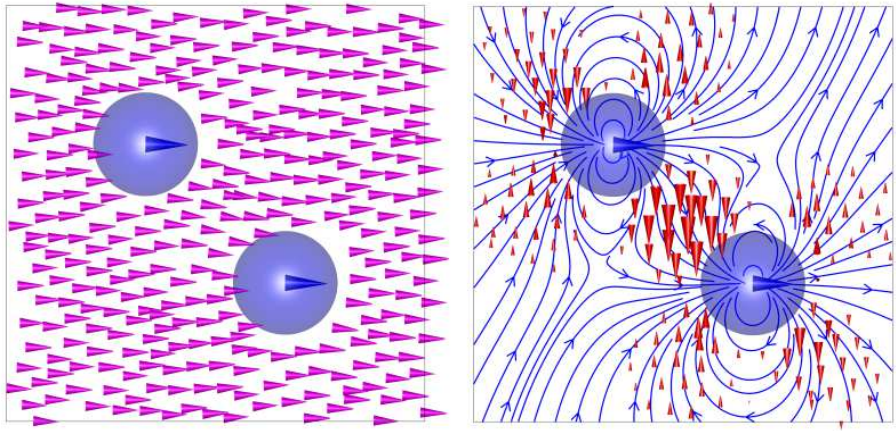}}
\caption{Results of a micromagnetic simulation for the 2D spin distribution around two selected nanoparticles (blue circles), which are assumed to be in a single-domain state. The external magnetic field $\mathbf{H}$ is applied horizontally in the plane ($\mu_0 H = 0.3 \, \mathrm{T}$). Left image: Magnetization distribution in both phases; note that $\mu_0 \Delta M = \mu_0 (M_{\mathrm{h}} - M_{\mathrm{s}}) = 1.5 \, \mathrm{T}$. In order to highlight the spin misalignment in the soft phase, the right image displays the magnetization component $\mathbf{M}_\perp$ perpendicular to $\mathbf{H}$ (red arrows). Thickness of arrows is proportional to the magnitude of $\mathbf{M}_\perp$. Blue lines: Dipolar field distribution.}
\label{fig13}
\end{figure*}

Generally, the sources of the magnetodipolar field are nonzero divergences of the magnetization ($\nabla \cdot \mathbf{M} \neq 0$). For magnetic nanocomposites, the most prominent ``magnetic volume charges'' are related to the nanoscale variations in the magnetic materials parameters at the phase boundary between particles and matrix, e.g., variations in the magnetization, anisotropy or exchange interaction. Such jumps in the magnetic materials parameters may give rise to an inhomogeneous spin structure which decorates each nanoparticle. Figure~\ref{fig13} displays the real-space magnetization distribution around two nanoparticles. Note that the symmetry of the spin structure replicates the symmetry of the $CT$ (compare to Fig.~\ref{fig6}). In the presence of an applied magnetic field the stray-field and associated magnetization configuration around each nanoparticle ``look'' similar (on the average), thus giving rise to \emph{dipolar correlations} which add up to a positive-definite $CT$ contribution to $d \Sigma_M / d \Omega$. Note, however, that for polycrystalline microstructures clover-leaf-type anisotropies may become only visible in $d \Sigma_M / d \Omega$ for $\mathbf{k}_0 \perp \mathbf{H}$.

As mentioned above, not only variations in the magnetization magnitude, but also variations in the \emph{direction} and/or \emph{magnitude} of the magnetic anisotropy $\bf{K}$ (random anisotropy) and variations in the magnitude of the exchange coupling may give rise to dipolar correlations. The micromagnetic simulation package allows us to vary the magnetic materials parameters of both phases of the nanocomposite. Hence, it becomes possible to study the impact of such situations on the magnetic SANS.

\begin{figure}[htb]
\centering
\resizebox{1.0\columnwidth}{!}{\includegraphics{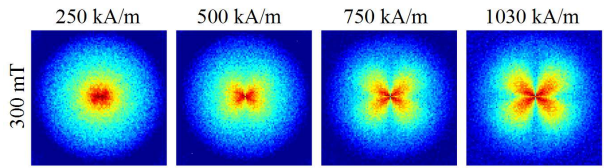}}
\caption{Fourier coefficient $|\widetilde{M}_y|^2(\mathbf{q})$ at $\mu_0 H = 0.3 \, \mathrm{T}$ and for $M_{\mathrm{h}} = M_{\mathrm{s}} = M$ (i.e., $\Delta M = 0$) ($\mathbf{k}_0 \perp \mathbf{H}$). $M$ increases from left to right (see insets). $K_{\mathrm{h}} = 4.6 \times 10^4 \, \mathrm{J/m}^3$, $K_{\mathrm{s}} = 1.0 \times 10^2 \, \mathrm{J/m}^3$ and random variations in easy-axis directions from particle to particle are assumed. Data taken from Ref.~\cite{michels2012prb1}.}
\label{fig14}
\end{figure}

In order to investigate variations in $\bf{K}$ (which are, by construction, naturally included into our micromagnetic algorithm), we have computed the spin distribution for the situation that $M_{\mathrm{h}} = M_{\mathrm{s}} = M$ (i.e., $\Delta M = 0$), but for different values of $M$. Figure~\ref{fig14} reveals that a clover-leaf-type pattern in $|\widetilde{M}_y|^2$ develops with increasing magnetization value $M$, i.e., with increasing strength of the magnetodipolar interaction. As jumps in the magnetization at phase boundaries are excluded here as possible sources for perturbations in the spin structure, it is straightforward to conclude that nanoscale fluctuations in $\bf{K}$ give rise to inhomogeneous magnetization states, which decorate each nanoparticle and which look similar to the structure shown in Fig.~\ref{fig13}. This observation strongly suggests that the origin of the clover-leaf pattern in $d \Sigma_M / d \Omega$ of nanomagnets is not only related to variations in magnetization magnitude but also due to variations in the magnitude and direction of the magnetic anisotropy field.

\section{Summary and conclusions}

By means of a recently developed micromagnetic simulation methodology---especially suited for modeling multi-phase materials---we have computed the magnetic small-angle neutron scattering (SANS) cross section $d \Sigma_M / d \Omega$ of a two-phase nanocomposite magnet from the NANOPERM family of alloys. Besides taking into account the full nonlinearity of Brown's equations of micromagnetics, the approach allows one to study the dependency of the \emph{individual} magnetization Fourier coefficients $\widetilde{M}_{(x,y,z)}$ on the applied magnetic field $\mathbf{H}$ and, most importantly, on the momentum-transfer vector $\mathbf{q}$. This ideally complements neutron experiments, in which a weighted sum of the $\widetilde{M}_{(x,y,z)}$ is generally measured. It is this particular circumstance, in conjunction with the flexibility of our micromagnetic package in terms of microstructure variation (particle size and distribution, materials parameters, texture, etc.), which makes us believe that the approach of combining full-scale 3D micromagnetic simulations with experimental magnetic-field-dependent SANS data will provide fundamental insights into the magnetic SANS of a wide range of magnetic materials. The micromagnetic simulations underline the importance of the magnetodipolar interaction for understanding magnetic SANS. In particular, the so-called clover-leaf-shaped angular anisotropy in $d \Sigma_M / d \Omega$---which was previously believed to be exclusively related to nanoscale jumps in the magnetization magnitude at internal interphases---is of relevance for all bulk nanomagnets with spatially fluctuating magnetic parameters.

\section*{Acknowledgments}

We thank the Deutsche Forschungsgemeinschaft (Project No.~BE 2464/10-1) and the National Research Fund of Luxembourg (ATTRACT Project No.~FNR/A09/01 and Project No.~FNR/10/AM2c/39) for financial support. Critical reading of the manuscript by Jens-Peter Bick and Dirk Honecker is gratefully acknowledged.

\bibliographystyle{elsarticle-num}

\begin{thebibliography}{10}
\expandafter\ifx\csname url\endcsname\relax
  \def\url#1{\texttt{#1}}\fi
\expandafter\ifx\csname urlprefix\endcsname\relax\def\urlprefix{URL }\fi
\expandafter\ifx\csname href\endcsname\relax
  \def\href#1#2{#2} \def\path#1{#1}\fi

\bibitem{fournet}
A.~Guinier, G.~Fournet, Small-Angle Scattering of X-rays, Wiley, New York,
  1955.

\bibitem{glatter}
O.~Glatter, O.~Kratky~(editors), Small-Angle X-ray Scattering, Academic Press,
  London, 1982.

\bibitem{feigin}
L.~A. Feigin, D.~I. Svergun, Structure Analysis by Small-Angle X-Ray and
  Neutron Scattering, Plenum Press, New York, 1987.

\bibitem{linchen}
S.-H. Chen, T.-L. Lin, in: D.~L. Price, K.~Sk\"{o}ld (Eds.), Methods of
  Experimental Physics--Neutron Scattering, Vol. 23-Part~B, Academic Press, San
  Diego, 1987, pp. 489--543.

\bibitem{zemb}
P.~Lindner, T.~Zemb~(editors), Neutron, X-Ray and Light Scattering:
  Introduction to an Investigative Tool for Colloidal and Polymeric Systems,
  North Holland, Amsterdam, 1991.

\bibitem{pedersen02}
J.~S. Pedersen, in: P.~Lindner, T.~Zemb (Eds.), Neutrons, X-Rays and Light:
  Scattering Methods Applied to Soft Condensed Matter, Elsevier, Amsterdam,
  2002, pp. 391--420.

\bibitem{svergun03}
D.~I. Svergun, M.~H.~J. Koch, Rep. Prog. Phys. 66 (2003) 1735.

\bibitem{stuhrmann04}
H.~B. Stuhrmann, Rep. Prog. Phys. 67 (2004) 1073.

\bibitem{wignall07}
Y.~B. Melnichenko, G.~D. Wignall, J. Appl. Phys. 102 (2007) 021101.

\bibitem{heinemann2000}
A.~Heinemann, H.~Hermann, A.~Wiedenmann, N.~Mattern, K.~Wetzig, J. Appl. Cryst.
  33 (2000) 1386.

\bibitem{hermann2000b}
H.~Hermann, A.~Heinemann, N.~Mattern, A.~Wiedenmann, Europhys. Lett. 51 (2000)
  127.

\bibitem{michels03epl}
A.~Michels, R.~N. Viswanath, J.~Weissm\"uller, Europhys. Lett. 64 (2003) 43.

\bibitem{herr04pss}
A.~Grob, S.~Saranu, U.~Herr, A.~Michels, R.~N. Viswanath, J.~Weissm\"uller,
  Phys. Status Solidi~A 201 (2004) 3354.

\bibitem{kohlbrecher05}
W.~Wagner, J.~Kohlbrecher, in: Y.~Zhu (Ed.), Modern Techniques for
  Characterizing Magnetic Materials, Kluwer Academic Publishers, Boston, 2005,
  pp. 65--103.

\bibitem{michels05epl}
A.~Michels, C.~Vecchini, O.~Moze, K.~Suzuki, J.~M. Cadogan, P.~K. Pranzas,
  J.~Weissm\"uller, Europhys. Lett. 72 (2005) 249.

\bibitem{michels05apl}
C.~Vecchini, O.~Moze, K.~Suzuki, P.~K. Pranzas, J.~Weissm\"uller, A.~Michels,
  Appl. Phys. Lett. 87 (2005) 202509.

\bibitem{michels06prb}
A.~Michels, C.~Vecchini, O.~Moze, K.~Suzuki, P.~K. Pranzas, J.~Kohlbrecher,
  J.~Weissm\"uller, Phys. Rev. B 74 (2006) 134407.

\bibitem{michels2010epjb}
D.~Honecker, A.~Ferdinand, F.~D\"obrich, C.~D. Dewhurst, A.~Wiedenmann,
  C.~G\'{o}mez-Polo, K.~Suzuki, A.~Michels, Eur. Phys. J. B 76 (2010) 209--213.

\bibitem{michels2012prb2}
A.~Michels, D.~Honecker, F.~D\"obrich, C.~D. Dewhurst, K.~Suzuki, A.~Heinemann,
  Phys. Rev. B 85 (2012) 184417.

\bibitem{bracchi04}
A.~Bracchi, K.~Samwer, P.~Schaaf, J.~F. L\"offler, S.~Schneider, Mater. Sci.
  Eng. A 375--377 (2004) 1027.

\bibitem{garcia04}
E.~Garc{\'i}a-Matres, A.~Wiedenmann, G.~Kumar, J.~Eckert, H.~Hermann,
  L.~Schultz, Phys. B 350 (2004) e315.

\bibitem{cald05}
R.~Garc{\'i}a~Calder{\'o}n, L.~Fern{\'a}ndez~Barqu{\'i}n, S.~N. Kaul, J.~C.
  G{\'o}mez~Sal, P.~Gorria, J.~S. Pedersen, R.~K. Heenan, Phys. Rev. B 71
  (2005) 134413.

\bibitem{mergia08}
K.~Mergia, S.~Messoloras, J. Phys.: Condens. Matter 20 (2008) 104219.

\bibitem{michels00a}
A.~Michels, J.~Weissm\"uller, A.~Wiedenmann, J.~G. Barker, J. Appl. Phys. 87
  (2000) 5953.

\bibitem{michels00b}
A.~Michels, J.~Weissm\"uller, A.~Wiedenmann, J.~S. Pedersen, J.~G. Barker,
  Philos. Mag. Lett. 80 (2000) 785.

\bibitem{michels02a}
A.~Michels, J.~Weissm\"uller, U.~Erb, J.~G. Barker, Phys. Status Solidi A 189
  (2002) 509.

\bibitem{michels02c}
A.~Michels, J.~Weissm\"uller, R.~Birringer, Eur. Phys. J. B 29 (2002) 533.

\bibitem{michels03prl}
A.~Michels, R.~N. Viswanath, J.~G. Barker, R.~Birringer, J.~Weissm\"uller,
  Phys. Rev. Lett. 91 (2003) 267204.

\bibitem{weissm04a}
J.~Weissm\"uller, A.~Michels, D.~Michels, A.~Wiedenmann, C.~E. Krill~III, H.~M.
  Sauer, R.~Birringer, Phys. Rev. B 69 (2004) 054402.

\bibitem{loeff05}
J.~F. L\"{o}ffler, H.~B. Braun, W.~Wagner, G.~Kostorz, A.~Wiedenmann, Phys.
  Rev. B 71 (2005) 134410.

\bibitem{michels08rop}
A.~Michels, J.~Weissm\"uller, Rep. Prog. Phys. 71 (2008) 066501.

\bibitem{michels08epl}
A.~Michels, F.~D\"obrich, M.~Elmas, A.~Ferdinand, J.~Markmann, M.~Sharp,
  H.~Eckerlebe, J.~Kohlbrecher, R.~Birringer, EPL 81 (2008) 66003.

\bibitem{elmas09}
A.~Michels, M.~Elmas, F.~D\"obrich, M.~Ames, J.~Markmann, M.~Sharp,
  H.~Eckerlebe, J.~Kohlbrecher, R.~Birringer, EPL 85 (2009) 47003.

\bibitem{michels2011jpcm}
D.~Honecker, F.~D\"obrich, C.~D. Dewhurst, A.~Wiedenmann, A.~Michels, J. Phys.:
  Condens. Matter 23 (2011) 016003.

\bibitem{dobrichprb2012}
F.~D\"obrich, J.~Kohlbrecher, M.~Sharp, H.~Eckerlebe, R.~Birringer, A.~Michels,
  Phys. Rev. B 85 (2012) 094411.

\bibitem{michels06a}
B.~van~den Brandt, H.~Gl\"attli, I.~Grillo, P.~Hautle, H.~Jouve,
  J.~Kohlbrecher, J.~A. Konter, E.~Leymarie, S.~Mango, R.~P. May, A.~Michels,
  H.~B. Stuhrmann, O.~Zimmer, Eur. Phys. J. B 49 (2006) 157.

\bibitem{forgan06}
M.~Laver, E.~M. Forgan, S.~P. Brown, D.~Charalambous, D.~Fort, C.~Bowell,
  S.~Ramos, R.~J. Lycett, D.~K. Christen, J.~Kohlbrecher, C.~D. Dewhurst,
  R.~Cubitt, Phys. Rev. Lett. 96 (2006) 167002.

\bibitem{forgan2011}
M.~R. Eskildsen, E.~M. Forgan, H.~Kawano-Furukawa, Rep. Prog. Phys. 74 (2011)
  124504.

\bibitem{bischof07}
M.~Bischof, P.~Staron, A.~Michels, P.~Granitzer, K.~Rumpf, H.~Leitner,
  C.~Scheu, H.~Clemens, Acta mater. 55 (2007) 2637.

\bibitem{weissm08}
G.~Balaji, S.~Ghosh, F.~D\"obrich, H.~Eckerlebe, J.~Weissm\"uller, Phys. Rev.
  Lett. 100 (2008) 227202.

\bibitem{kreyssig09}
A.~Kreyssig, R.~Prozorov, C.~D. Dewhurst, P.~C. Canfield, R.~W. McCallum, A.~I.
  Goldman, Phys. Rev. Lett. 102 (2009) 047204.

\bibitem{gazeau02}
F.~Gazeau, E.~Dubois, J.-C. Bacri, F.~Bou{\'e}, A.~Cebers, R.~Perzynski, Phys.
  Rev. E 65 (2002) 031403.

\bibitem{thomson05}
T.~Thomson, S.~L. Lee, M.~F. Toney, C.~D. Dewhurst, F.~Y. Ogrin, C.~J. Oates,
  S.~Sun, Phys. Rev. B 72 (2005) 064441.

\bibitem{ijiri05}
Y.~Ijiri, C.~V. Kelly, J.~A. Borchers, J.~J. Rhyne, D.~F. Farrell, S.~A.
  Majetich, Appl. Phys. Lett. 86 (2005) 243102.

\bibitem{albi06}
A.~Wiedenmann, U.~Keiderling, K.~Habicht, M.~Russina, R.~G\"ahler, Phys. Rev.
  Lett. 97 (2006) 057202.

\bibitem{grigorieva07}
N.~A. Grigoryeva, S.~V. Grigoriev, H.~Eckerlebe, A.~A. Eliseev, A.~V. Lukashin,
  K.~S. Napolskii, J. Appl. Cryst. 40 (2007) s532.

\bibitem{napolski07}
K.~S. Napolskii, A.~A. Eliseev, N.~V. Yesin, A.~V. Lukashin, Y.~D. Tretyakov,
  N.~A. Grigorieva, S.~V. Grigoriev, H.~Eckerlebe, Physica E 37 (2007) 178.

\bibitem{bonini07}
M.~Bonini, A.~Wiedenmann, P.~Baglioni, J. Appl. Cryst. 40 (2007) s254.

\bibitem{oku09}
T.~Oku, T.~Kikuchi, T.~Shinohara, J.~Suzuki, Y.~Ishii, M.~Takeda, K.~Kakurai,
  Y.~Sasaki, M.~Kishimoto, M.~Yokoyama, Y.~Nishihara, Physica B 404 (2009)
  2575.

\bibitem{gri10}
S.~V. Grigoriev, A.~V. Syromyatnikov, A.~P. Chumakov, N.~A. Grigoryeva, K.~S.
  Napolskii, I.~V. Roslyakov, A.~A. Eliseev, A.~V. Petukhov, H.~Eckerlebe,
  Phys. Rev. B 81 (2010) 125405.

\bibitem{krycka2010}
K.~L. Krycka, R.~A. Booth, C.~R. Hogg, Y.~Ijiri, J.~A. Borchers, W.~C. Chen,
  S.~M. Watson, M.~Laver, T.~R. Gentile, L.~R. Dedon, S.~Harris, J.~J. Rhyne,
  S.~A. Majetich, Phys. Rev. Lett. 104 (2010) 207203.

\bibitem{disch2012}
S.~Disch, E.~Wetterskog, R.~P. Hermann, A.~Wiedenmann, U.~Vainio,
  G.~Salazar-Alvarez, L.~Bergstr\"om, T.~Br\"uckel, New J. Phys. 14 (2012)
  013025.

\bibitem{laver2010}
M.~Laver, C.~Mudivarthi, J.~R. Cullen, A.~B. Flatau, W.-C. Chen, S.~M. Watson,
  M.~Wuttig, Phys. Rev. Lett. 105 (2010) 027202.

\bibitem{uehland2010}
B.~G. Ueland, J.~W. Lynn, M.~Laver, Y.~J. Choi, S.-W. Cheong, Phys. Rev. Lett.
  104 (2010) 147204.

\bibitem{lister2010}
S.~J. Lister, T.~Thomson, J.~Kohlbrecher, K.~Takano, V.~Venkataramana, S.~J.
  Ray, M.~P. Wismayer, M.~A. de~Vries, H.~Do, Y.~Ikeda, S.~L. Lee, Appl. Phys.
  Lett. 97 (2010) 112503.

\bibitem{dufour2011}
C.~Dufour, M.~R. Fitzsimmons, J.~A. Borchers, M.~Laver, K.~L. Krycka,
  K.~Dumesnil, S.~M. Watson, W.~C. Chen, J.~Won, S.~Singh, Phys. Rev. B 84
  (2011) 064420.

\bibitem{pappas09}
C.~Pappas, E.~Leli$\mathrm{\grave{e}}$vre-Berna, P.~Falus, P.~M. Bentley,
  E.~Moskvin, S.~Grigoriev, P.~Fouquet, B.~Farago, Phys. Rev. Lett. 102 (2009)
  197202.

\bibitem{gri09}
S.~V. Grigoriev, D.~Chernyshov, V.~A. Dyadkin, V.~Dmitriev, S.~V. Maleyev,
  E.~V. Moskvin, D.~Menzel, J.~Schoenes, H.~Eckerlebe, Phys. Rev. Lett. 102
  (2009) 037204.

\bibitem{pflei2009}
S.~M\"uhlbauer, B.~Binz, F.~Jonietz, C.~Pfleiderer, A.~Rosch, A.~Neubauer,
  R.~Georgii, P.~B\"oni, Science 323 (2009) 915.

\bibitem{brown}
W.~F. Brown~Jr., Micromagnetics, Interscience Publishers, New York, 1963.

\bibitem{aharonibook}
A.~Aharoni, Introduction to the Theory of Ferromagnetism, 2nd Edition,
  Clarendon Press, Oxford, 1996.

\bibitem{kronfahn03}
H.~Kronm\"uller, M.~F\"ahnle, Micromagnetism and the Microstructure of
  Ferromagnetic Solids, Cambridge University Press, Cambridge, 2003.

\bibitem{kron63}
H.~Kronm\"uller, A.~Seeger, M.~Wilkens, Z. Phys. 171 (1963) 291.

\bibitem{suzuki06}
K.~Suzuki, G.~Herzer, in: D.~Sellmyer, R.~Skomski (Eds.), Advanced Magnetic
  Nanostructures, Springer, New York, 2006, pp. 365--401.

\bibitem{kronparkinhandbook07}
H.~Kronm\"uller, S.~Parkin, Handbook of Magnetism and Advanced Magnetic
  Materials, John Wiley \& Sons, Chichester, 2007, vol.2, pp.795.

\bibitem{fidler05}
J.~Fidler, P.~Speckmayer, T.~Schrefl, D.~Suess, J. Appl. Phys. 97 (2005)
  10E508.

\bibitem{ogrin06}
F.~Y. Ogrin, S.~L. Lee, M.~Wismayer, T.~Thomson, C.~D. Dewhurst, R.~Cubitt,
  S.~M. Weekes, J. Appl. Phys. 99 (2006) 08G912.

\bibitem{herr08pss}
S.~Saranu, A.~Grob, J.~Weissm\"uller, U.~Herr, Phys. Status Solidi A 205 (2008)
  1774.

\bibitem{oommf}
M.~Donahue, D.~Porter, \href{http://math.nist.gov/oommf/}{The Object Oriented
  MicroMagnetic Framework (OOMMF)}.
\newline\urlprefix\url{http://math.nist.gov/oommf/}

\bibitem{jodrey85}
W.~S. Jodrey, E.~M. Tory, Phys. Rev. A 32 (1985) 2347.

\bibitem{gutfleisch2006}
O.~Gutfleisch, K.-H. M\"uller, K.~Khlopkov, M.~Wolf, A.~Yan, R.~Sch\"afer,
  T.~Gemming, L.~Schultz, Acta Mater. 54 (2006) 997.

\bibitem{gutfleisch2010}
G.~Hrkac, T.~G. Woodcock, C.~Freeman, A.~Goncharov, J.~Dean, T.~Schrefl,
  O.~Gutfleisch, Appl. Phys. Lett. 97 (2010) 232511.

\bibitem{berkov05handbook}
D.~V. Berkov, N.~L. Gorn, in Handbook of Advanced Magnetic Materials, Vol.~2,
  Springer Verlag and Tsinghua University Press, 2005, pp. 421--507, edited by
  Y. Liu, D. Shindo and D. J. Sellmyer.

\bibitem{erokhin2011ieee}
S.~Erokhin, D.~Berkov, N.~Gorn, A.~Michels, IEEE Trans. Magn. 47 (2011) 3044.

\bibitem{gibbon02}
P.~Gibbon, G.~Sutmann, in Quantum Simulations of Complex Many-Body Systems:
  From Theory to Algorithms, Lecture Notes, edited by J. Grotendorst, D. Marx
  and A. Muramatsu 10 (2002) 467--506.

\bibitem{berkov98}
D.~V. Berkov, N.~L. Gorn, Phys. Rev. B 57 (1998) 14332.

\bibitem{gorn07}
N.~L. Gorn, D.~V. Berkov, P.~G{\"o}rnert, D.~Stock, J. Magn. Magn. Mat. 310
  (2007) 2829.

\bibitem{hockneybook}
R.~W. Hockney, J.~W. Eastwood, {Computer simulation using particles},
  McGraw-Hill, 1981.

\bibitem{deserno98}
M.~Deserno, C.~Holm, J. Chem. Phys. 109 (1998) 7678.

\bibitem{landau35}
L.~Landau, E.~Lifshitz, Phys. Z. Sowjetunion 8 (1935) 153.

\bibitem{micromagus}
D.~V. Berkov, N.~L. Gorn, \href{http://www.micromagus.de}{{Micromagus: package
  for micromagnetic simulations}}.
\newline\urlprefix\url{http://www.micromagus.de}

\bibitem{erokhin2012prb}
S.~Erokhin, D.~Berkov, N.~Gorn, A.~Michels, Phys. Rev. B 85 (2012) 024410.

\bibitem{rave98}
W.~Rave, K.~Fabian, A.~Hubert, J. Magn. Magn. Mater. 190 (1998) 332.

\bibitem{albi2010}
A.~Wiedenmann, Collection SFN 11 (2010) 219, http://www.neutron-sciences.org/.

\bibitem{michels2012prb1}
S.~Erokhin, D.~Berkov, N.~Gorn, A.~Michels, Phys. Rev. B 85 (2012) 134418.

\end{thebibliography}

\end{document}